\documentclass[useAMS,usenatbib]{mn2e}
\usepackage{amsmath,amsfonts,epsfig,natbib}
\def\be{\begin{equation}} 
\def\ee{\end{equation}} 
\def\ba{\begin{eqnarray}} 
\def\ea{\end{eqnarray}}

\def\reff@jnl#1{{\rm#1\/}}

\def\aj{\reff@jnl{AJ}}                  
\def\araa{\reff@jnl{ARA\&A}}            
\def\apj{\reff@jnl{ApJ}}                        
\def\apjl{\reff@jnl{ApJ}}               
\def\apjs{\reff@jnl{ApJS}}              
\def\ao{\reff@jnl{Appl.Optics}}         
\def\apss{\reff@jnl{Ap\&SS}}            
\def\aap{\reff@jnl{A\&A}}               
\def\aapr{\reff@jnl{A\&A~Rev.}}         
\def\aaps{\reff@jnl{A\&AS}}             
\def\azh{\reff@jnl{AZh}}                        
\def\baas{\reff@jnl{BAAS}}              
\def\jrasc{\reff@jnl{JRASC}}            
\def\memras{\reff@jnl{MmRAS}}           
\def\mnras{\reff@jnl{MNRAS}}            
\def\pra{\reff@jnl{Phys.Rev.A}}         
\def\prb{\reff@jnl{Phys.Rev.B}}         
\def\prc{\reff@jnl{Phys.Rev.C}}         
\def\prd{\reff@jnl{Phys.Rev.D}}         
\def\prl{\reff@jnl{Phys.Rev.Lett}}      
\def\pasp{\reff@jnl{PASP}}              
\def\pasj{\reff@jnl{PASJ}}              
\def\qjras{\reff@jnl{QJRAS}}            
\def\skytel{\reff@jnl{S\&T}}            
\def\solphys{\reff@jnl{Solar~Phys.}}    
\def\sovast{\reff@jnl{Soviet~Ast.}}     
\def\ssr{\reff@jnl{Space~Sci.Rev.}}     
\def\zap{\reff@jnl{ZAp}}                        
\def\nat{\reff@jnl{Nature}}             

\title[Searching for non-Gaussianity in the VSA data]{Searching for
non-Gaussianity in the VSA data}

\author[Richard Savage et al.]
{Richard Savage$^1$, 
 Richard A. Battye$^2$,
 Pedro Carreira$^2$, 
 Kieran Cleary$^2$, 
 Rod D. Davies$^2$, 
\newauthor 
 Richard J. Davis$^2$,  
 Clive Dickinson$^2$, 
 Ricardo Genova-Santos$^3$,
 Keith Grainge$^1$, 
\newauthor
 Carlos M. Guti{\'e}rrez$^3$,  
 Yaser A. Hafez$^2$,
 Michael P. Hobson$^1$,  
 Michael E. Jones$^1$, 
\newauthor
 R\"udiger Kneissl$^1$, 
 Katy Lancaster$^1$, 
 Anthony Lasenby$^1$,  
 J. P. Leahy$^2$,
\newauthor 
 Klaus Maisinger$^1$,  
 Guy G. Pooley$^1$, 
 Nutan Rajguru$^1$, 
 Rafael Rebolo$^{3,4}$, 
 Graca Rocha$^{1,5}$, 
\newauthor 
 Jos\'e Alberto Rubi\~no-Martin$^{3, \ddagger}$,  
 Pedro Sosa Molina$^3$,  
 Richard D.E. Saunders$^1$, 
 Paul Scott$^1$,  
\newauthor 
 An\v ze Slosar$^1$, 
 Angela C. Taylor$^1$,  
 David Titterington$^1$,  
 Elizabeth Waldram$^1$, 
\newauthor 
 Robert A. Watson$^{2,\dagger}$
\\
  $^1$ Astrophysics Group, Cavendish Laboratory, University of Cambridge, UK\\
  $^2$ University of Manchester, Jodrell Bank Observatory, UK\\
  $^3$ Instituto de Astrof{\'i}sica de Canarias, 38200 La Laguna,
  Tenerife, Spain.\\
  $^4$Consejo Superior de Investigaciones Cient{\'{\i}}ficas, Spain \\
  $^5$Centro de Astrof\'{\i}sica da Universidade do Porto, R. das
Estrelas s/n, 4150-762 Porto, Portugal \\
  $^{\dagger}$Present address: Instituto de Astrof{\'{\i}}sica de
Canarias.\\
  $^{\ddagger}$Present address: Max-Planck Institut f\"ur Astrophysik,
  Garching, Germany}

  \pagerange{\pageref{firstpage}--\pageref{lastpage}}

\pubyear{2003}

\begin{document}
\maketitle
\label{firstpage}
\begin{abstract}
We have tested Very Small Array (VSA) observations of three regions of
sky for  the presence of non-Gaussianity,  using high-order cumulants,
Minkowski functionals, a wavelet-based test and a Bayesian joint power
spectrum/non-Gaussianity analysis.  We find  the data from two regions
to be consistent  with Gaussianity.  In the third  region, we obtain a
96.7\% detection of non-Gaussianity using the wavelet test. We perform
simulations to characterise the tests,  and conclude that this is
consistent with expected residual point source contamination.  There
is therefore no evidence  that   this  detection  is  of   cosmological  origin.   Our
simulations show  that the tests  would be sensitive to any  residual point
sources above the data's source subtraction level of 20~mJy.  The tests
are also sensitive to cosmic string networks at an rms fluctuation
level of  $105 ~ \mu K$ (i.e. equivalent to the best-fit observed value).
They are  not sensitive to string-induced fluctuations if an equal rms
of Gaussian CDM fluctuations 
is added, thereby reducing the fluctuations due to the strings network
to $74~\mu K$ rms .  We  especially  highlight   the  usefulness  of
non-Gaussianity testing in eliminating systematic effects from our
data.
\end{abstract}


\section{Introduction}
The search for non-Gaussianity in the anisotropies of the cosmic
microwave background (CMB) radiation is of major importance to modern
cosmology.  Different fundamental theories of structure formation
predict markedly different non-Gaussian CMB signatures, making the
quantification of any such signatures a powerful discriminator between
these possibilities.  Measurements of the CMB angular power spectrum
have now ruled out the classic non-Gaussian example of topological
defects as the primary method of generating CMB anisotropy, although
the possibility of a sub-dominant contribution remains.  
Recent work on different varieties of the inflationary paradigm
\citep[e.g.][]{Peebles-ng-inflation-1999,Wang-inflation-bispectrum-2000,
Bartolo-liddle-curvaton-2002} has shown that different forms of
inflation can also be distinguished by the characterisation of any
non-Gaussianity present in the CMB.

There are other very good reasons to search for non-Gaussianity in the
CMB.  Much of CMB cosmology centres on the accurate determination of
the angular power spectrum.  If the CMB fluctuations form a Gaussian
random field, then the power spectrum completely characterises its
statistical properties.  Therefore, the detection of any
non-Gaussianity would indicate that there is further information to be
gathered.  It is also the case that, with one exception
\citep{Graca-joint-estimation-2001}, all power spectrum analyses of
CMB data assume the data to be Gaussian.  As with any assumption, this
must be tested if we are to have confidence in the results.

There is also the consideration of contamination of CMB data sets.
Extragalactic sources, Sunyaev-Z'eldovich decrements and the various
Galactic foregrounds all play their part in contaminating the CMB.
Moreover, they all introduce non-Gaussianity into the data, making
non-Gaussianity testing an excellent way to identify their presence.
Furthermore, data contamination can occur other than by the telescope
seeing something unwanted in the sky.  Systematic effects are
invariably an issue.  The detection of non-Gaussianity in the COBE
data by \citet{Ferreira-COBE-ng-detection-1998} was traced to such an
effect \citep{COBE-systematic} and, in general, non-Gaussianity
testing can be extremely useful in identifying and removing
systematics.

As a consequence of the many-fold importance of non-Gaussianity
testing to CMB cosmology, many different techniques have been
developed to investigate it \citep[see
e.g.][]{Ferreira-cumulants-1997,
Hobson-wavelet-ng-test,Graca-joint-estimation-2001,Verde-trispectrum-2001,
Hansen-harmonic-ng-test-2002,Chiang-phase-ng-2002,
Aghanim-wavelet-fourier-ng-2003}. A number of techniques have also been
applied to the data from various CMB instruments over the last few
years \citep[see
e.g.][]{Wu-maxima-ng-2001,Polenta-boomerang-ng-2002,
Chiang-wmap-ng-2003,komatsu-wmap-ng-testing-2003,
DeTroia-boomerang-trispectrum-2003, Santos-maxima-bispectrum-2003}
These analyses have shown the practical application of non-Gaussianity
testing, particularly the importance of robust Monte Carlo simulation
for determining the significance level of any result.

In this paper we present the results of applying a selection of
non-Gaussianity tests to the VSA data set first presented in its
entirety in \citet{VSApaperV} (Paper V hereafter) and
\citet{VSApaperVI} (Paper VI hereafter).  Results from the VSA in its
compact configuration have already been presented in
\citet{VSApaperI}, \citet{VSApaperII}, \citet{VSApaperIII}, and
\citet{VSApaperIV}, (hereafter Papers I - IV respectively).

\section{Interferometric observations of the CMB}
\begin{figure*}
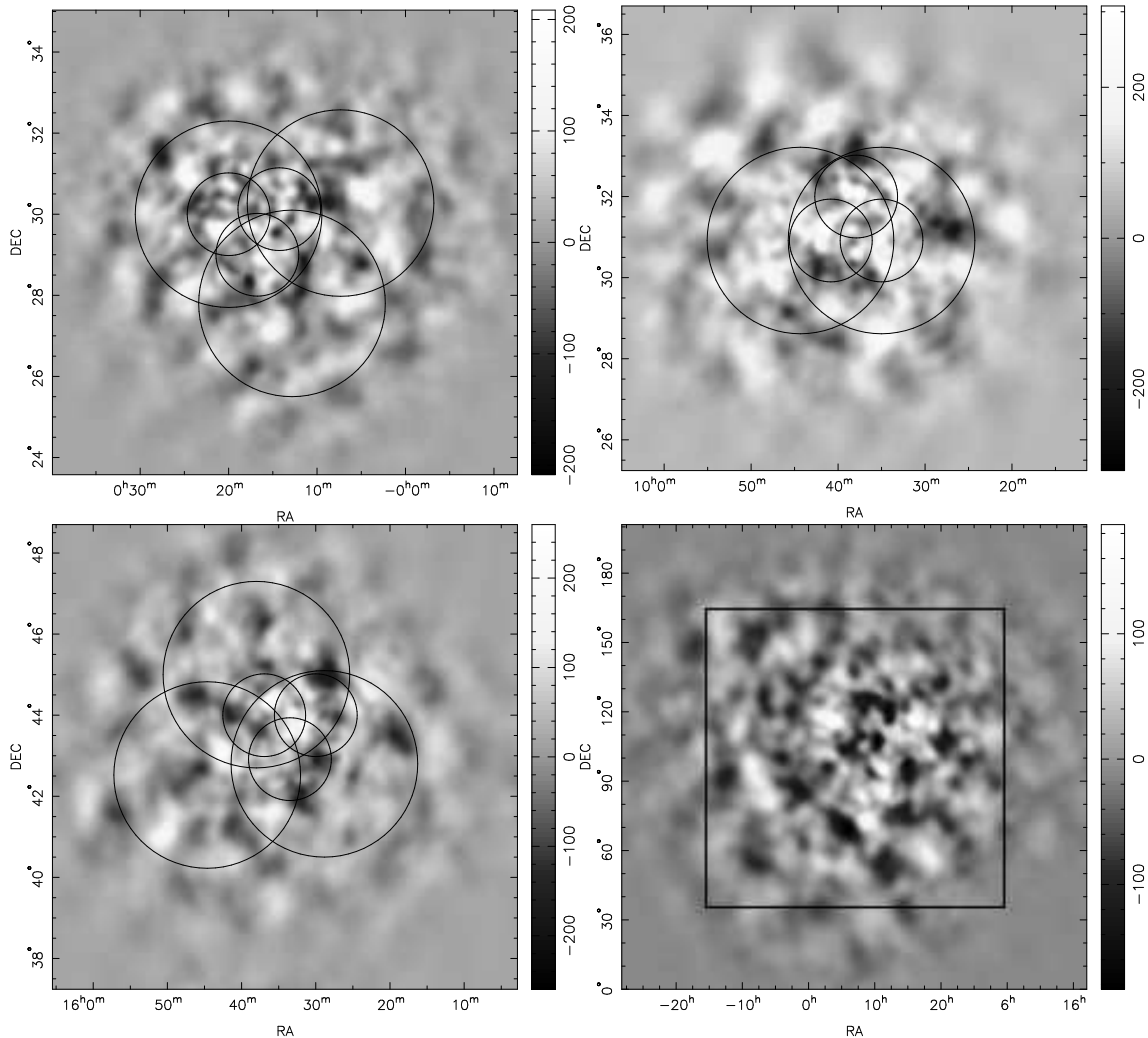

\begin{minipage}{150mm}
\epsfig{file=VSA1mosaic.ps , angle=-90, width=7.5cm}
\epsfig{file=VSA2mosaic.ps , angle=-90, width=7.5cm}
\epsfig{file=VSA3mosaic.ps , angle=-90, width=7.5cm}
\epsfig{file=region.ps, angle=-90, width=7.5cm}
\caption{Top left, top right and bottom left: The VSA1, VSA2 and VSA3
mosaics.  The maps are produced using a MEM algorithm.  The circles
denote the FWHM of the primary beams of each of the individual pointings.
Bottom right:  a simulated Gaussian CMB realisation corresponding to
the VSA1 mosaic.  The box shows
the central 128$\times$128 pixel region that is analysed in each case.  The box is
chosen to enclose the FWHMs but exclude the area of low sensitivity
towards the edge of the whole map.  In all cases, a central region
128x128 pixels, corresponding to $7.36^\circ \times 7.36^\circ$.}
\label{fig:maps}
\end{minipage}
\end{figure*}

\subsection{The VSA observations}
The VSA is a 14-element interferometric telescope operating with a
1.5~GHz observing bandwidth, centred at 34~GHz for the data analysed
in this paper.  It has been used in two distinct configurations, the
compact array (covering the multipole range $\ell \sim150$ -- $900$
with a 34~GHz primary beam FWHM of $4.6^{\circ}$) and the extended
array (extending to multipoles as high as $\ell
\sim1400$ with a 34~GHz primary beam FWHM of $2.0^{\circ}$).

The VSA observing strategy has focussed so far on making deep
observations on 3 relatively small patches of sky (named the VSA1,
VSA2 and VSA3 mosaics).  These cover a total of 101 square degrees of
sky, with  each region comprising the data from either five or six
separate pointing centres (three extended array fields and either two
or three compact array fields).  These individual fields are
coherently mosaiced to produce maps and power spectra.

Using a combination of a 15~GHz blind survey \citep{waldram-9C-survey-2003}
and regular monitoring observations  made
at 34~GHz by the VSA's dedicated source subtraction
interferometer, we have subtracted all sources with flux densities
$S_{34~GHz} > 20$~mJy (or, in the case of the compact array fields, 80~mJy) from our data.

The data set analysed in this paper is described more fully in Paper V.

\subsection{The VSA maps}
Some of our chosen tests operate on CMB maps produced from our data.
We use a maximum-entropy method (MEM) based on that of
\citet{Maisinger-MEM-1997}
to produce mosaiced maps of the VSA data. Maisinger et al. show that
this method produces more accurate reconstructions than the
simpler Wiener filter approach.  Furthermore, it has been shown 
\citep{Maisinger-MEM-topol-1998}
that this method accurately reconstructs non-Gaussian features 
in CMB maps, such as the prominent hot spots produced by cosmic
strings.  This is not true of Wiener filtering.

The MEM maps of the three VSA mosaics are shown in Fig.
\ref{fig:maps}.  Well outside the FWHM of the primary beams the
data contain negligible useful information about the CMB structure.
Therefore, the MEM maps
show no features towards their edges.  As these featureless
regions contain no useful CMB information, we select only the central
region of our maps for analysis.  In all cases these regions are
128$\times$128 pixels, corresponding to $7.36^\circ \times
7.36^\circ$.  An example of this is shown in Fig. \ref{fig:maps} (bottom
right).  It can be seen that the region size corresponds approximately
to the combined area within the FWHM of the primary beams.  

It is noted that this choice of pixel size significantly oversamples the VSA data
set considered here.  For example, the compact array synthesised beam
has a FWHM of approximately 30'.  These maps therefore contain of
order a few hundred independent elements each.

\subsection{Particular issues in detecting non-Gaussianity in
interferometer data}
There are a number of characteristics of interferometer observations
that require careful consideration when testing for non-Gaussianity.

First, an interferometer does not measure the brightness
distribution of the sky directly.  Rather, it samples (regions of)
the visibility plane, which is the Fourier transform of the sky
brightness distribution convolved with the aperture illumination
function (itself the Fourier transform of the primary beam).  It is 
therefore most natural to test for non-Gaussianity in the visibility
plane, where receiver noise is Gaussian and uncorrelated between
samples, and the signal is uncorrelated on scales larger than the
aperture illumination function.  The interferometer's
intrinsic sensitivity to specific Fourier modes can also be
exploited, for example, in searching for a non-Gaussian signature
dominant only on certain angular scales.  Targeting these scales
essentially eliminates the contribution from non-overlapping
Gaussian CMB fluctuations on other scales, which would otherwise mask
the non-Gaussian signature. 

By contrast,{\it images} produced from interferometer data have a
number of features which tend to complicate any statistical analysis
performed on them.

\begin{itemize}
\item{The noise contributions between the individual pixels of an interferometer map
are correlated over a wide range of angular scales.}
\item{Interferometers sample the Fourier plane
incompletely.  This leads to interferometer maps being convolved by a
so-called 'dirty beam', with its
associated sidelobes.  This effect is not very
significant for the VSA, as it has good sampling coverage of the
visibility plane.  It is also noted that the MEM algorithm we employ
attempts to deconvolve the maps it produces.}
\item{The combination of data obtained using different primary beams
and array configurations (as in the case of the VSA) leads to significant
variation in temperature sensitivity and angular resolution across the
map.  This means that the effective probability density function (PDF)
from which the map pixels are drawn will be non-stationary.}  
\end{itemize}
Nevertheless, all of these complications can be resolved by the
use of appropriate Monte Carlo simulations. 

%
%

\section{Choice of non-Gaussianity test}
There are an infinite number of ways for a PDF to be non-Gaussian
but any given non-Gaussianity analysis is sensitive only to
a subset of these.  It is therefore prudent to apply a number of
different analyses.  In the following subsections, we describe the analyses
employed in this paper.

For the sake of clarity, in this paper we adopt the following
terminology.  We employ four different non-Gaussianity analyses.  Each
of these analyses consists of a number of individual statistics, for
which we assess significance levels.  For the map plane analyses, we
then form a number of different combined tests.  Each test consists of a
set of individual statistics.  A test function is calculated from this
set of statistics and used to assess a combined significance level.

\subsection{Analysis in the visibility plane}
We perform a Bayesian joint estimation of non-Gaussianity
and the power spectrum of CMB anisotropies, before marginalising over
the power spectrum to give an estimate of the non-Gaussianity of the
data.  We use the general form of
the likelihood,  derived from the eigenfunctions of a linear harmonic
oscillator, as developed by 
\citet{Graca-joint-estimation-2001}.
This distribution takes the form of a Gaussian multiplied by the square of a
(finite) series of Hermite polynomials, where the coefficients 
$\alpha_{n}$ are used as non-Gaussian estimators.
These amplitudes $\alpha_{n}$ can be written as series of
cumulants \citep{Contaldi-99}.  In particular, they
can be set to zero independently without mathematical inconsistency.
Furthermore, perturbatively (that is when the cumulants are ``small''
in a suitable sense), the amplitudes $\alpha_n$ are proportional to 
the $n^{th}$ order cumulant. We modify the standard 
Gaussian maximum likelihood method for
analysing interferometer observations \citep{madcow} with our new
distribution.  Instead of assuming the simple Gaussian
form for the probability distribution of each signal-to-noise
eigenmode $\xi_i$ \citep{bond-98}, 
we consider the more general situation in which all
$\alpha_n$ are set to zero, except for the real part of
one of them. 
To illustrate, we consider the $\alpha_3$ likelihood function
\be\label{like}
P(x)= \frac{e^{-\frac{x^{2}}{2 \sigma_{0}^{2}}}}{\sqrt{2 \pi}
\sigma_{0}} 
\left[ \alpha_{0} + \frac{\alpha_{3}}{\sqrt{48}} 
H_{3} \left(\frac{x}{\sqrt{2} \sigma_{0}}\right) \right]^{2}, 
\ee
with $\alpha_0= \sqrt{1- \alpha_3^{2}}$ to ensure that $P(x)$ is
normalised to unity.

We approximate the generalisation of this distribution to the multidimensional
case, in the signal-to-noise basis, by simply taking the product of the
individual one-dimensional distributions.
Here the data are the signal-to-noise eigenmodes, $\xi_i$, which are
uncorrelated and have a covariance matrix given by
$\langle \xi_i \xi_j \rangle = (1+a_k\lambda_i)\delta_{ij}$, 
where $\lambda_i$ is the eigenvalue corresponding to the eigenmode $\xi_i$ and 
$a_k = \langle \ell^2 C_\ell/(2\pi) \rangle_{\mbox{{\small $k$}}}$, i.e. $a_k$ is the average of value of $\ell^2 C_\ell/(2\pi)$ in the
$k$th spectral bin.  

\begin{figure*}
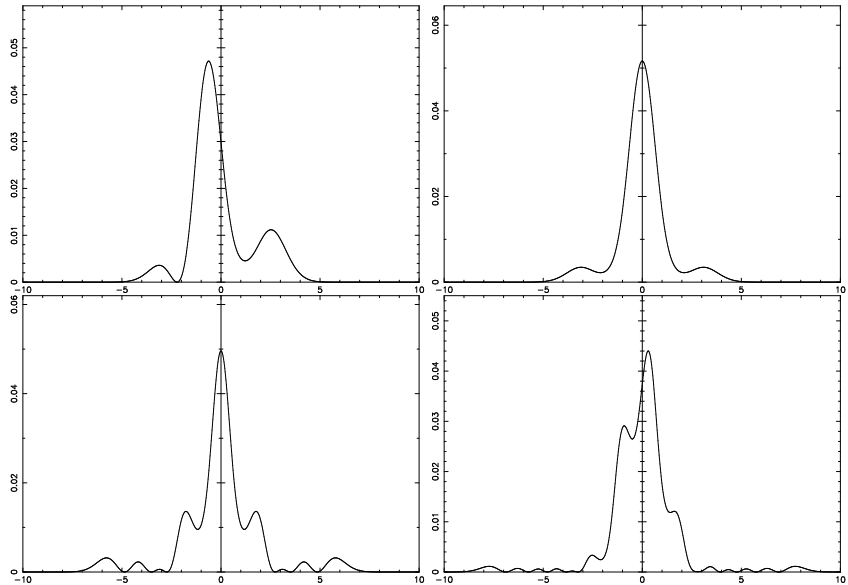

\begin{minipage}{150mm}
\begin{center}
\epsfig{file=hermiteN3.ps , angle=-90, width=5.5cm}
\epsfig{file=hermiteN4.ps , angle=-90, width=5.5cm}
\epsfig{file=hermiteN10.ps , angle=-90, width=5.5cm}
\epsfig{file=hermiteN17.ps , angle=-90, width=5.5cm}
\caption{Some fairly extreme examples of the non-Gaussian PDFs used in the joint estimation test.
Shown are PDFs with non-Gaussian parameter values of $a_3 = 0.5$ (top left), $a_4
= 0.3$ (top right), $a_{10} = -0.4$ (bottom left) and $a_{17} = 0.25$
(bottom right).}
\label{fig:ng-pdf}
\end{center}
\end{minipage}
\end{figure*}

Thus, when considering the power spectrum in the $k$th spectral bin,
we adopt the likelihood function
\be
{\cal L}(\xi|a_k,\alpha_3)
=\prod_{i} \frac{e^{-\frac{\xi_{i}^{2}}{2c_i}}}   
{\sqrt{2\pi c_i}} 
\left[ \alpha_{0} + \frac{\alpha_{3}}{\sqrt{48}} 
H_{3}\left(\frac{\xi_{i}}{\sqrt{2 (c_i)}}\right)\right]^{2}, 
\ee
where $c_i = 1+a_k\lambda_i$.
The $\alpha_3$ could in principle depend
on $\ell$, but for simplicity we have dropped this dependence.
The cases of this test considered in this paper are for a single
non-zero  $\alpha_{n}$ with $n = 3-20$.
Some examples of such non-Gaussian PDFs are shown in Fig.
\ref{fig:ng-pdf}.

It is important to note that this method does not ensure that the
higher order moments are uncorrelated and so this approximation will
be invalid in the strong non-Gaussian limit.  Such a case, however,
would be straightforward to detect using other methods and this is
thus not a significant problem.

We note that the phase mapping technique of 
\citet{Chiang-phase-ng-2002}
is also well suited to analysing interferometer data.  For
our data set, however, the individual visibility points have small
signal-to-noise ratios.  For example, each point on the VSA power spectrum 
(see Paper V) is the result of a likelihood analysis of several 
thousand binned visibility points.  Therefore, such a phase mapping
technique would tend to measure information about the VSA's noise 
properties rather than the CMB.  

\subsection{Analysis in the map plane}
For analysis of possible non-Gaussianity in the map plane, we
calculate three distinct sets of statistics.  
\begin{description}
\item{\it Map cumulants}.  The higher-order ($\kappa_3$ and above) cumulants of
a Gaussian distribution are all zero,   
\citep[see e.g.][]{Ferreira-cumulants-1997}
therefore, the cumulants of a set of data can conveniently be used to
determine whether that data set is drawn from a Gaussian
distribution or not.  We estimate the higher-order ($\kappa_3-\kappa_8$) cumulants of
the pixels in our maps.  
\item{\it Minkowski functionals}.  For a
Euclidean 2-D image such as our maps, the three Minkowski functionals
are the surface area, perimeter and Euler characteristic of an
excursion region \citep[see e.g.][]{Hobson-wavelet-ng-test}.  This
region is taken as the part of the map above
a certain threshold temperature, $T$.  The functionals  are therefore
functions of this variable.  We calculate the 3 Minkowski functionals 
of our maps at 51 different value of $T$ between $\pm 6~ \sigma$ (where
$\sigma$ is the rms of the map pixels).  The individual statistics are
therefore the functional values at each of these 51 values of $T$.
\item{\it Wavelet cumulants}.  This analysis is not actually calculated in the map
plane, but the map provides the starting point.  Following the method of 
\citet{Hobson-wavelet-ng-test}
we perform a wavelet transform on our map. We then estimate the
skewness and kurtosis of subsets of the wavelet co-efficients using
$k$-statistics.  There are a total of 21 different subsets, corresponding
to wavelets of given physical 
scales.  This makes the wavelet test ideal for detecting a given
non-Gaussian feature repeated many times in a map.  The individual
statistics are therefore the 21 separate skewness and 21 separate kurtosis
values.  These are
repeated for each of the nine wavelet bases considered by Hobson et
al, giving $9 \times 42$ wavelet statistics in total.
\end{description}

\subsection{Evaluation of individual confidence levels}
The joint estimation analysis uses a Bayesian
framework and so calculates a full marginalised posterior probability
distribution of the non-Gaussian parameter for each statistic.  For the map plane
statistics, however, we adopt a frequentist approach and so must define the Gaussian
confidence regions on our statistics in terms of appropriately simulated observations of
Gaussian sky realisations. To do this, we simulate a large number
(1000 for the results given in this paper) of equivalent Gaussian
realisations (EGR), calculate the statistical values for each one, and
use these values to define the Gaussian confidence regions on each
individual statistic.  This also allows us to account for the complications of
analysing interferometer maps.

For the purposes of this paper, we produce our EGR as follows:

\begin{itemize}
\item{The real and imaginary parts of each Fourier mode
$a_{lm}$ are drawn independently from a Gaussian random distribution. 
The variance of this PDF is determined by an underlying power
spectrum, given by a CDM model using the
VSA best-fit parameters for this data set (see Paper VI).}
\item{The modes are inverse Fourier transformed to produce a Gaussian CMB
realisation.}
\item{For each field in the simulated mosaic, the
realisation is multiplied by the appropriate primary beam for that
field, and Fourier transformed to produce the visibility
planes for each field.}
\item{The visibility planes are sampled using a real VSA observation as a
template.  Gaussian noise is then added to each sampled
visibility, using the same rms as given in the VSA template file.}
\item{A map is then made from this simulated mosaiced observation,
using the same software pipeline that is used to produce the real
maps.}
\end{itemize}

The EGR are therefore simulated observations of a
Gaussian sky with a realistic power spectrum, sampling the visibility
plane correctly for a given observation and with correct noise
contributions.

\subsection{Evaluation of combined confidence levels}
\begin{figure}
\epsfig{file=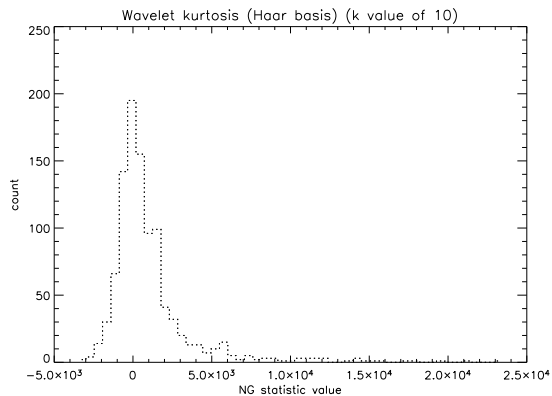 , angle=0, width=7.5cm}
\caption{Example of a non-normally distributed histogram of
statistical values.  Shown is the histogram for the kurtosis
wavelet test applied to 1000 EGR, using the Haar wavelet basis.  
}
\label{fig:non-normal-histogram}
\end{figure}
\begin{figure*}
\begin{minipage}{150mm}
\epsfig{file=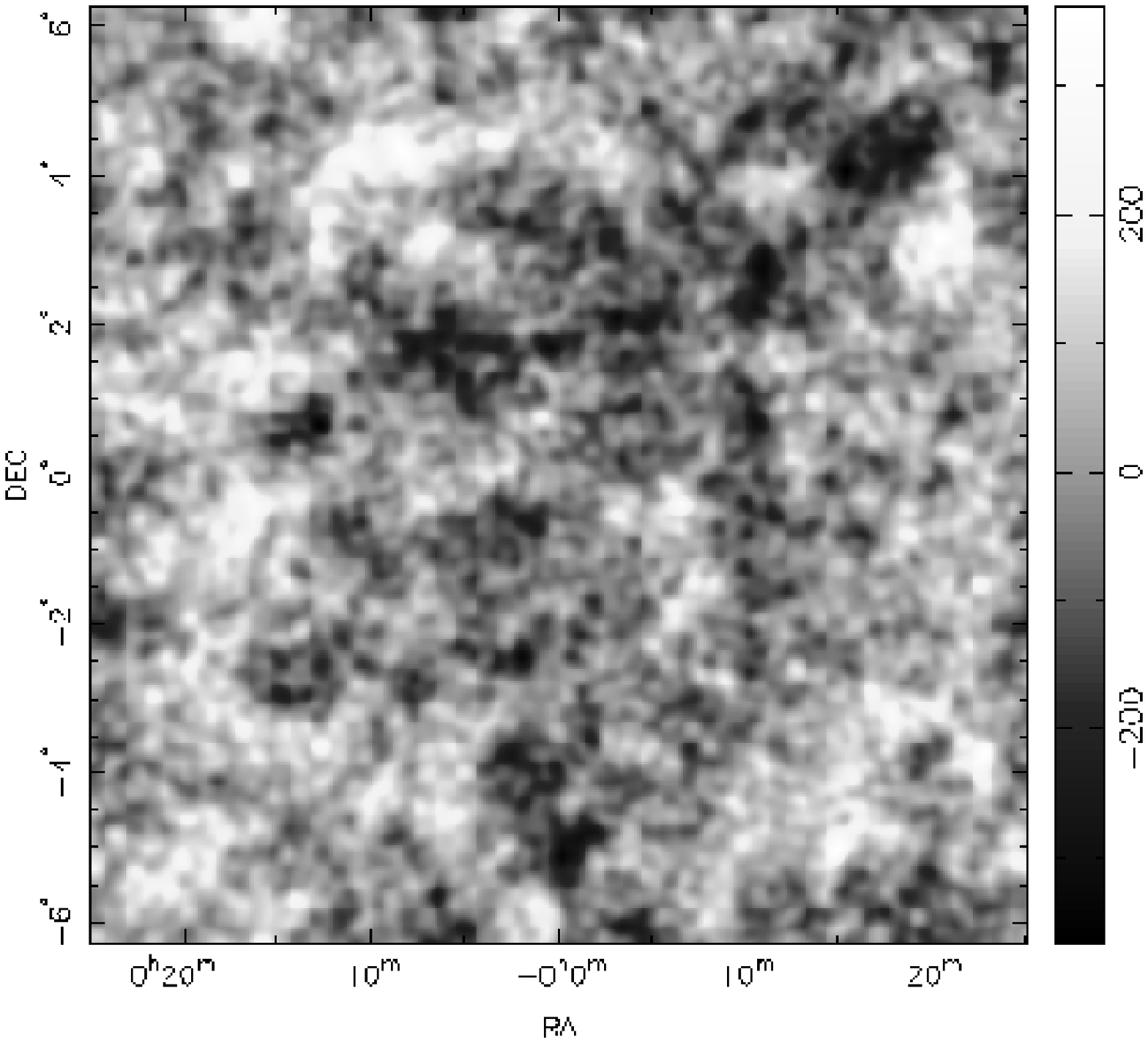 , angle=-90, width=5cm}
\epsfig{file=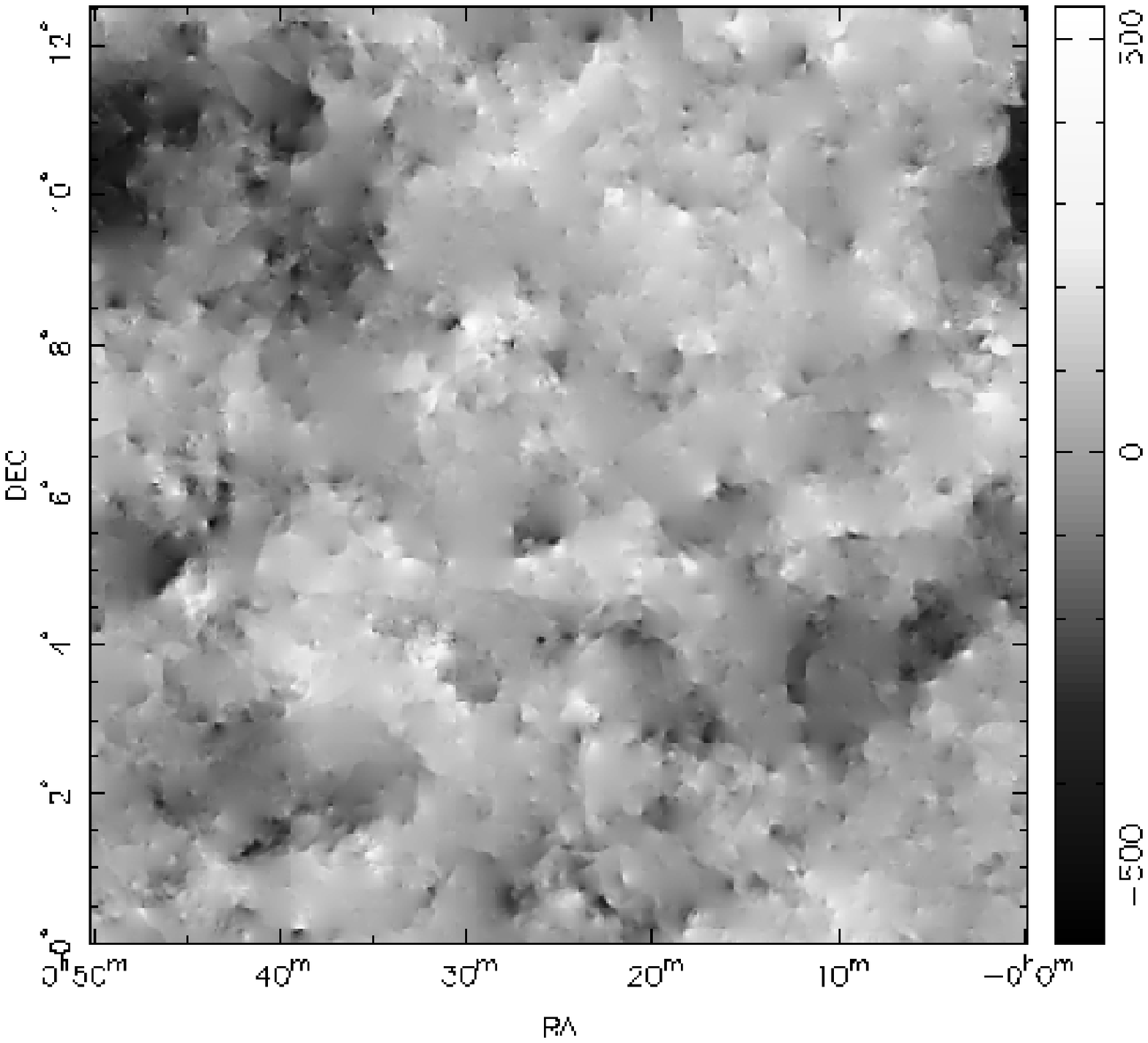 , angle=-90, width=5cm}
\epsfig{file=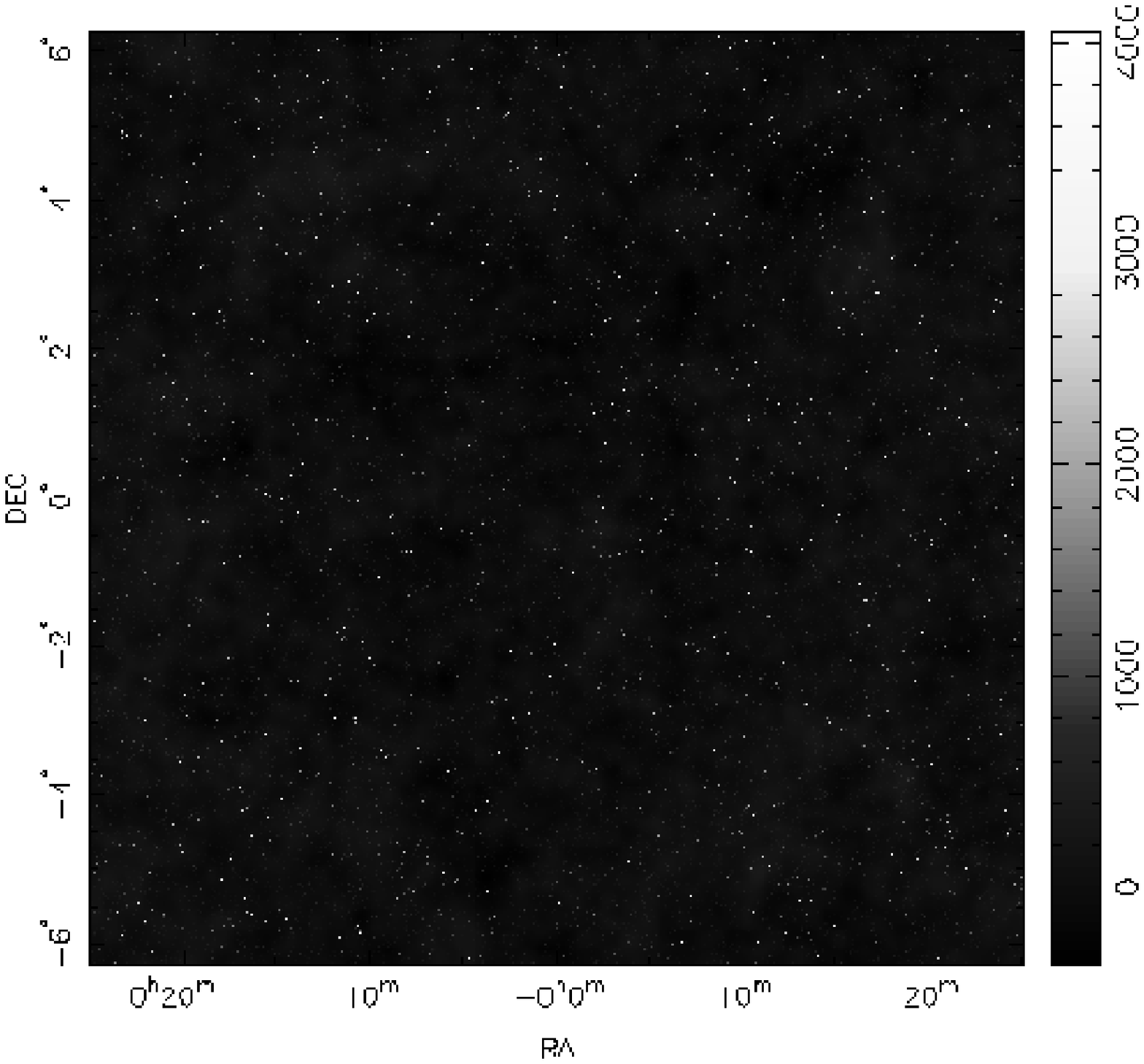 , angle=-90, width=5cm}
\caption{Three of the simulated-skies used in the simulations.  Shown are the
Gaussian CMB realisation (left), the pure cosmic strings map (middle) and the
CMB/point sources (subtracted to 5mJy) (right).}
\label{fig:true_skies}
\end{minipage}
\end{figure*}
We have detailed methods of obtaining significance levels for all
of the
individual statistics we are considering.  However, we can place
stronger limits on non-Gaussianity by combining these statistics in
varying ways.  We note that this will
not be attempted for the Bayesian joint estimation analysis, which
assumes a different PDF model in each case, and also
that the CMB power spectrum is divided into different band powers,
making such a combination unsuitable.  We therefore only combine the
map plane statistics.  The combinations used are listed in the
first column of Table \ref{tab:map_results}.

To produce an overall significance level for a given combination, we need to
calculate the value of some test function of the individual statistics
included.  An 
obvious choice would be a chi-squared test.  However, many of the
statistics used in this paper are drawn from non-normal distributions
(see e.g. Fig. \ref{fig:non-normal-histogram}),
making a standard chi-squared test unsuitable.  A non-normal version of the
chi-squared has been derived \citep{Ferreira-COBE-ng-detection-1998}
but as noted by \citet{Barreiro-01}
for the wavelet statistics in particular, it is very hard to evaluate.  We
therefore follow the method of \citet{Mukherjee-00} and use a test
function that is simply the number of individual statistics in the
combination that show a 95\% significant deviation from Gaussianity.  The
combined significance level is therefore the proportion of EGRs with
test function values smaller than that obtained for the real data.
It is noted that this method will not identify a single highly
non-Gaussian statistic value.  However, any such case will be self
evident, and so this does not present a problem.

\section{Analysis of simulated observations}
In order to characterise the effectiveness of our chosen
non-Gaussianity analyses in detecting various types of non-Gaussianity,
we applied them to simulated
VSA observations of a number of underlying simulated-skies.  The
simulated observations are produced in the same way as the EGR (see
above), except that the underlying Fourier modes are provided by the
Fourier transform of the simulated-sky. The simulated observations
have the same relative pointing centres, visibility-plane sampling and
rms noise levels as the six fields (three compact array and three
extended array) comprising the VSA1 mosaic. Therefore, the simulations
realistically sample the visibility plane and include realistic levels
of Gaussian noise.

We considered the following simulated-skies (see Fig.
\ref{fig:true_skies}):

\begin{description}
\item{\bf Gaussian CMB realisation.}  
A Gaussian realisation of the
CMB with a power spectrum given by the VSA best-fit cosmological
parameters for the CDM cosmological model (see Paper VI).
This is included principally to ensure that our non-Gaussianity tests
do not produce spurious detections.
\item{\bf Pure cosmic strings}. 
A map containing cosmic strings networks \citep{bouchet-strings-1988}.
The map is scaled to have a pixel rms equal to that of the Gaussian
CMB map.  It is noted that we only use a single realisation here.
While it is clearly preferable to run any analysis on many different
realisations, the lack of a large set of strings
simulations prevents this.  The analyses of this map do not therefore contain
the effect of sample variance. 
\item{\bf Composite Gaussian/strings.}  
The sum of the above Gaussian CMB
and strings maps, rescaled to have the same rms as either individual map.
\item{\bf Bright point sources/CMB.}  
A power law distribution of point sources defined by the extrapolation
to 34~GHz of 15~GHz source counts made by the Ryle telescope (see
Paper II for details).
\item{\bf Source subtracted.}  
The same point source distribution plus
CMB as above, but with all sources above a given flux density removed
(to mimic the VSA source subtraction strategy).  Subtraction levels of
20mJy, 10mJy and 5mJy are considered.  The VSA data analysed in this
paper has a nominal source subtraction level of 20mJy.  See Paper II
for a more detailed discussion.
\end{description}
\subsection{Visibility plane simulations}
\begin{figure}
\epsfig{file=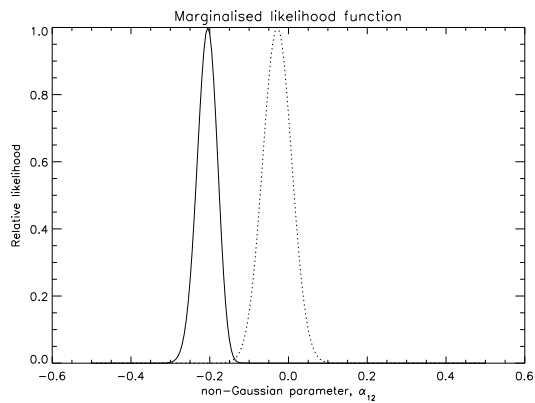 , angle=0, width=7.5cm}
\caption{Example detection of the presence of bright points sources in
the data using the joint estimation analysis.  Shown is the marginalised posterior probability
distribution calculated from simulated extended array data in an
annular visibility plane band centred about $\ell = 1085$.  The solid
and dashed lines show the results with and without the bright point sources.}
\label{fig:hermite_point_source}
\end{figure}
\begin{figure*}
\begin{minipage}{150mm}
\epsfig{file=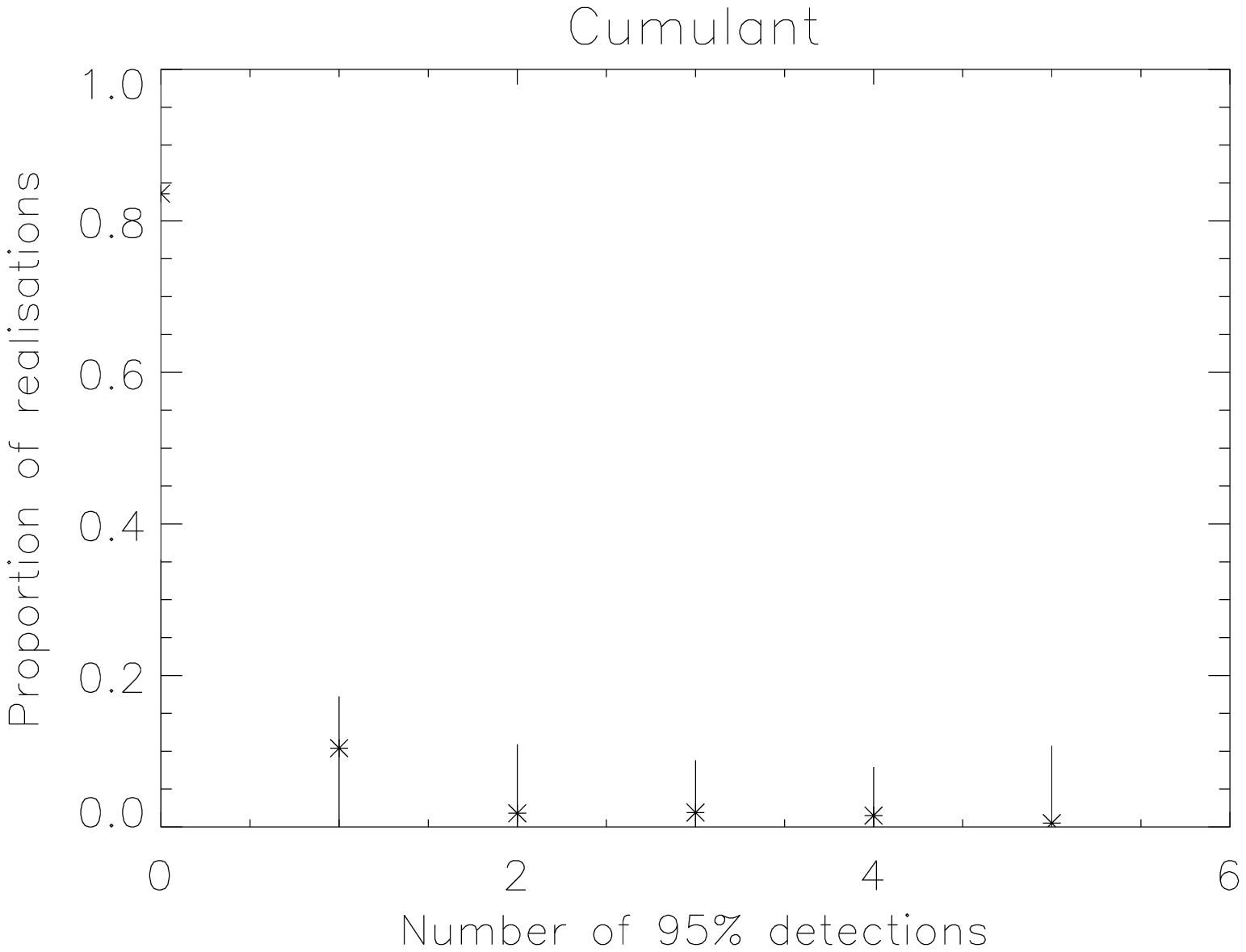 , angle=0, width=5cm}
\epsfig{file=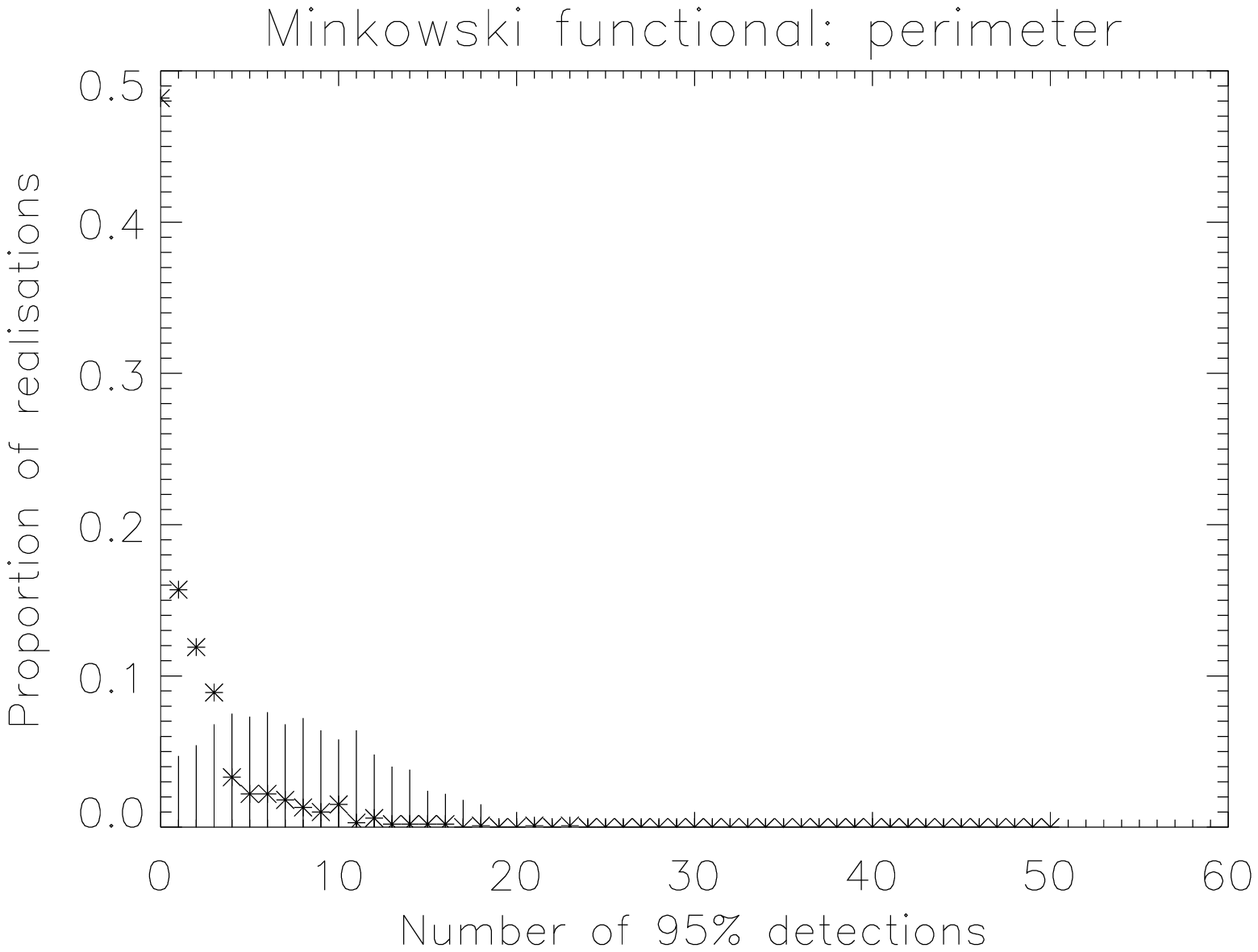 , angle=0, width=5cm}
\epsfig{file=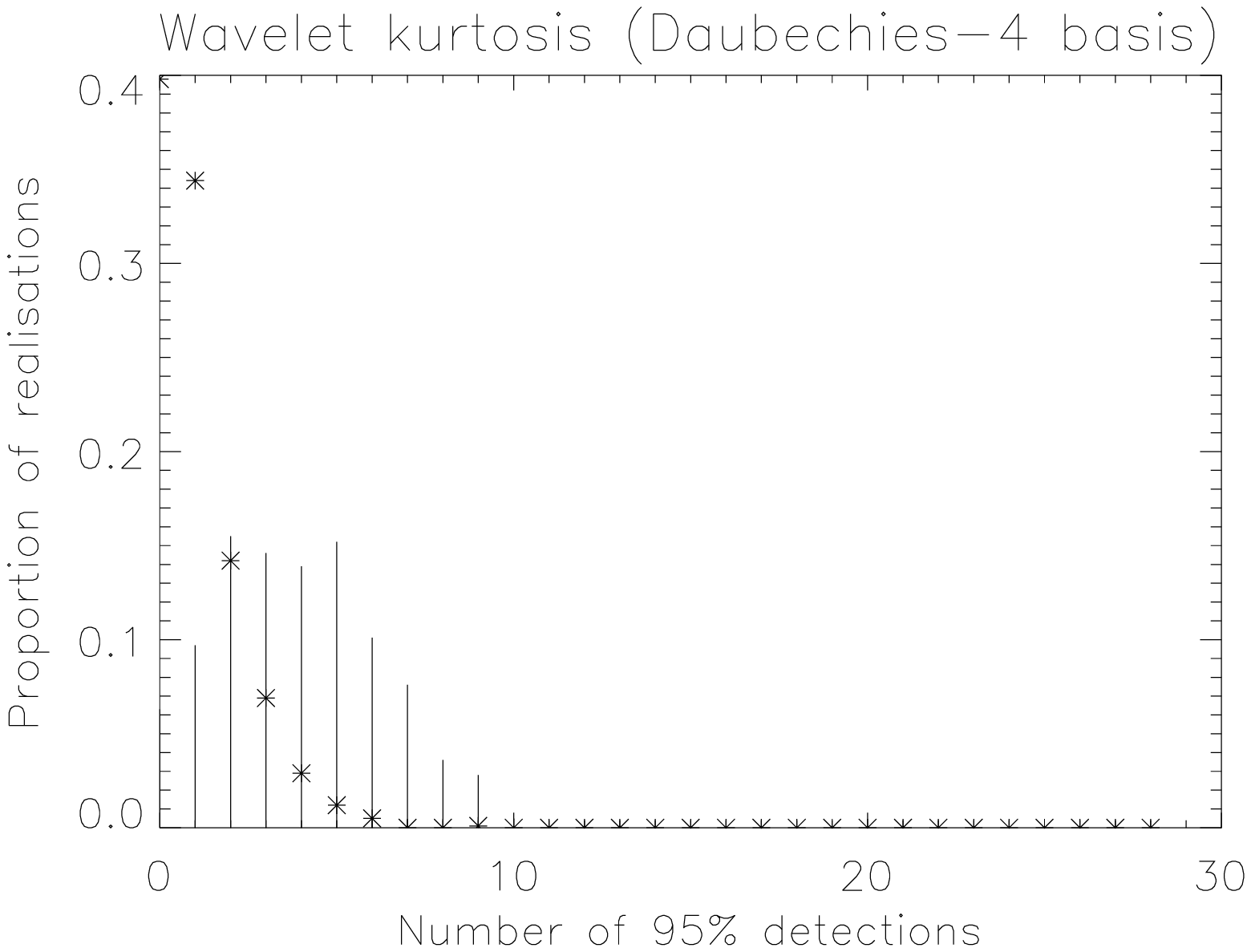 , angle=0, width=5cm}
\end{minipage}
\begin{minipage}{150mm}
\epsfig{file=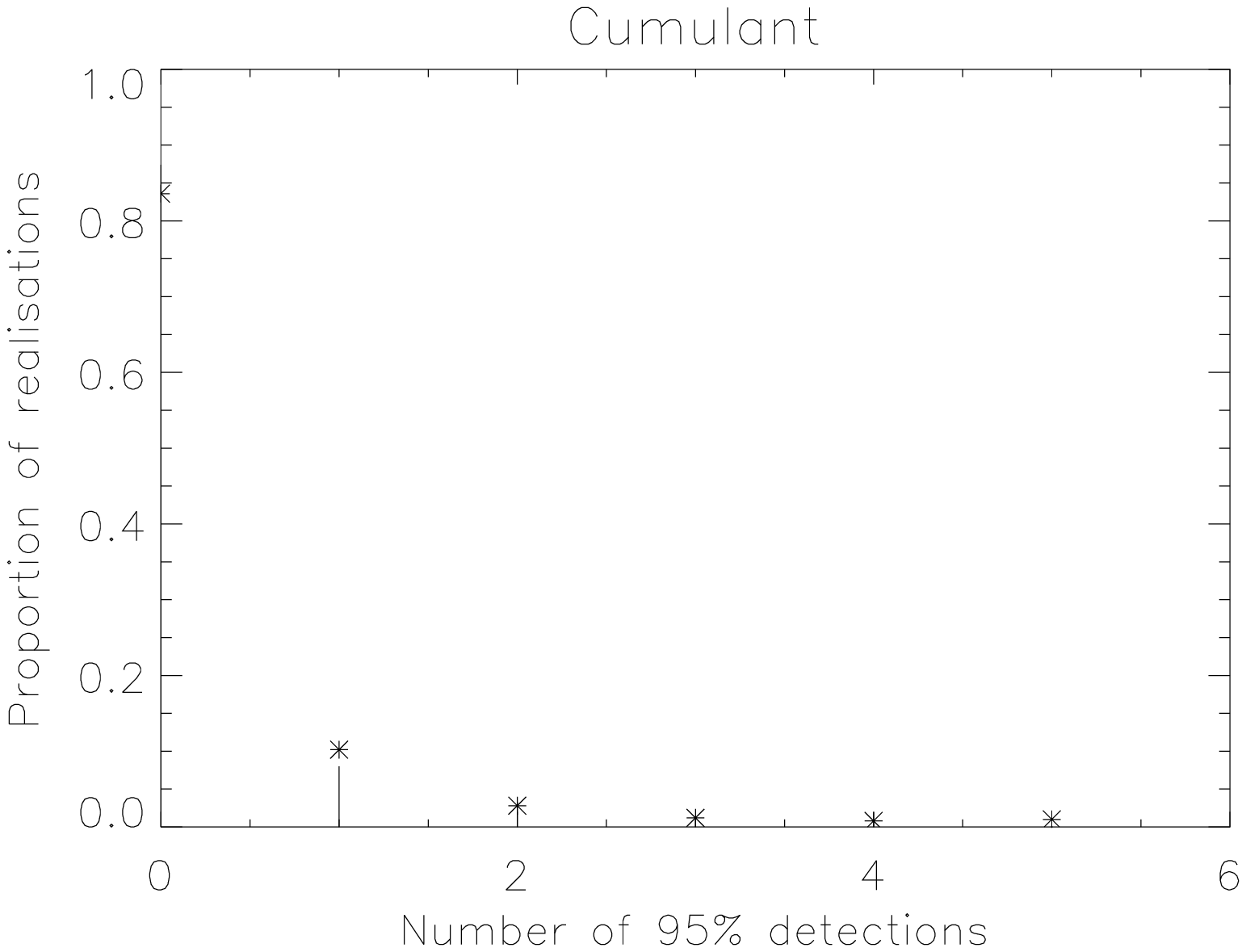 , angle=0, width=5cm}
\epsfig{file=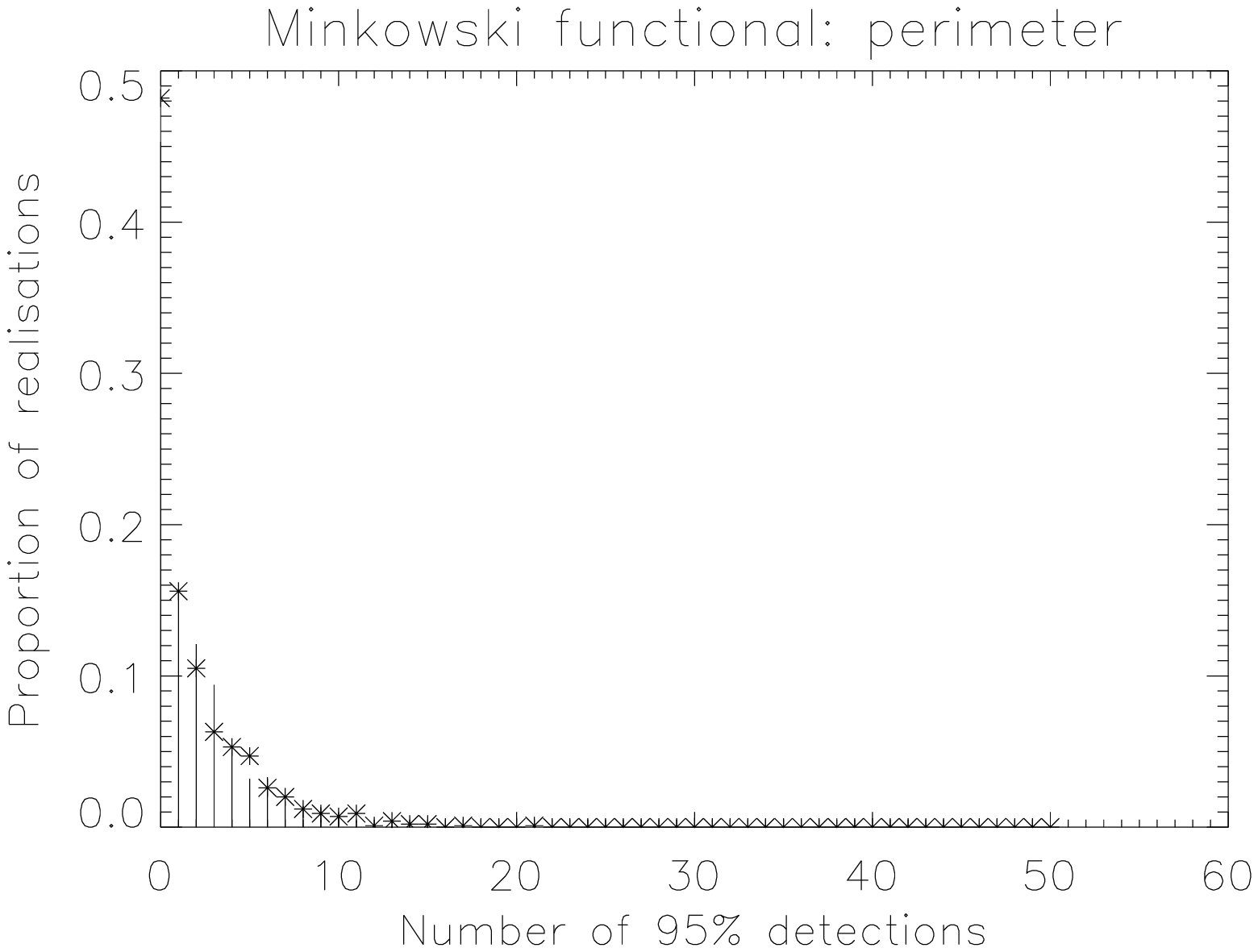 , angle=0, width=5cm}
\epsfig{file=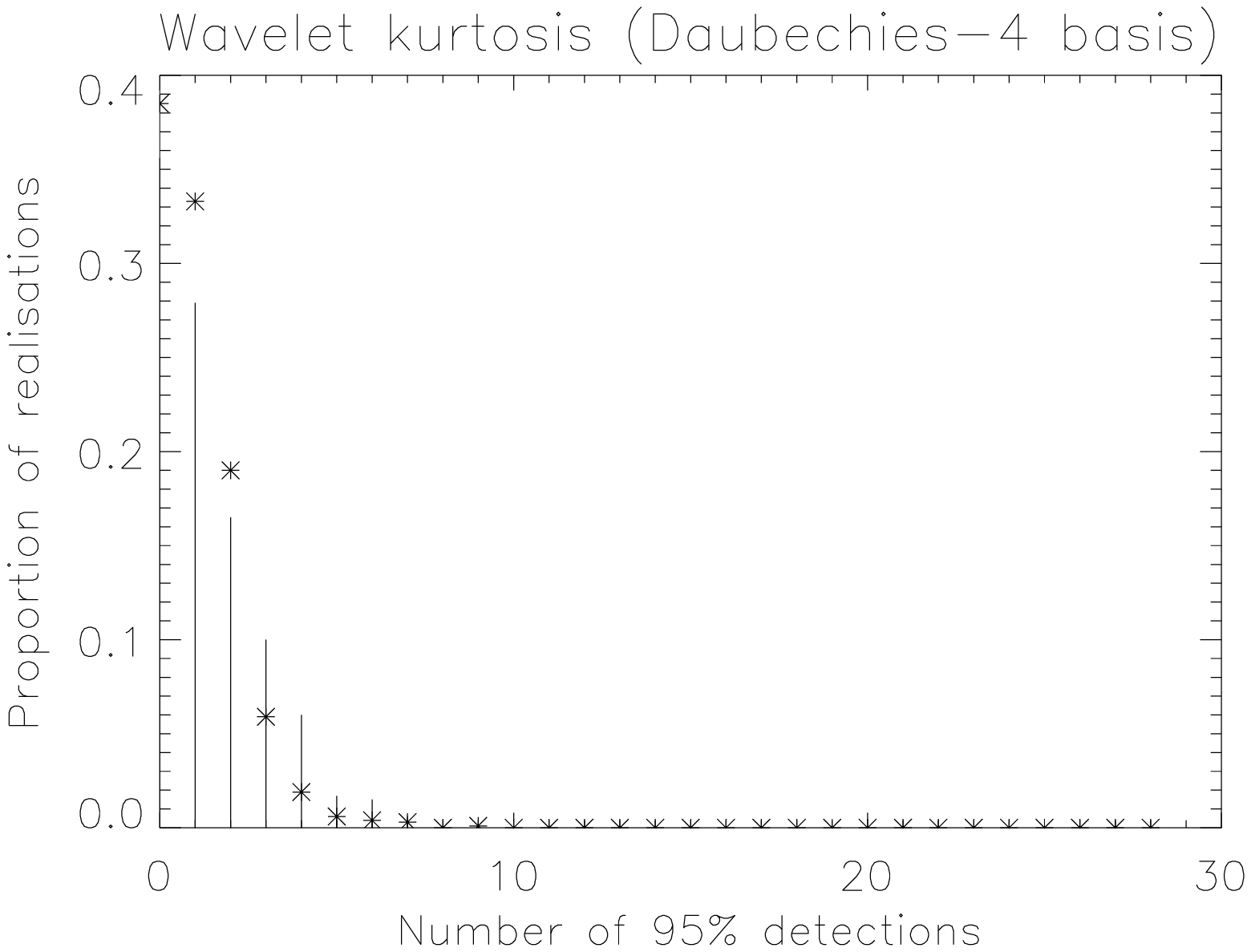 , angle=0, width=5cm}
\end{minipage}
\begin{minipage}{150mm}
\epsfig{file=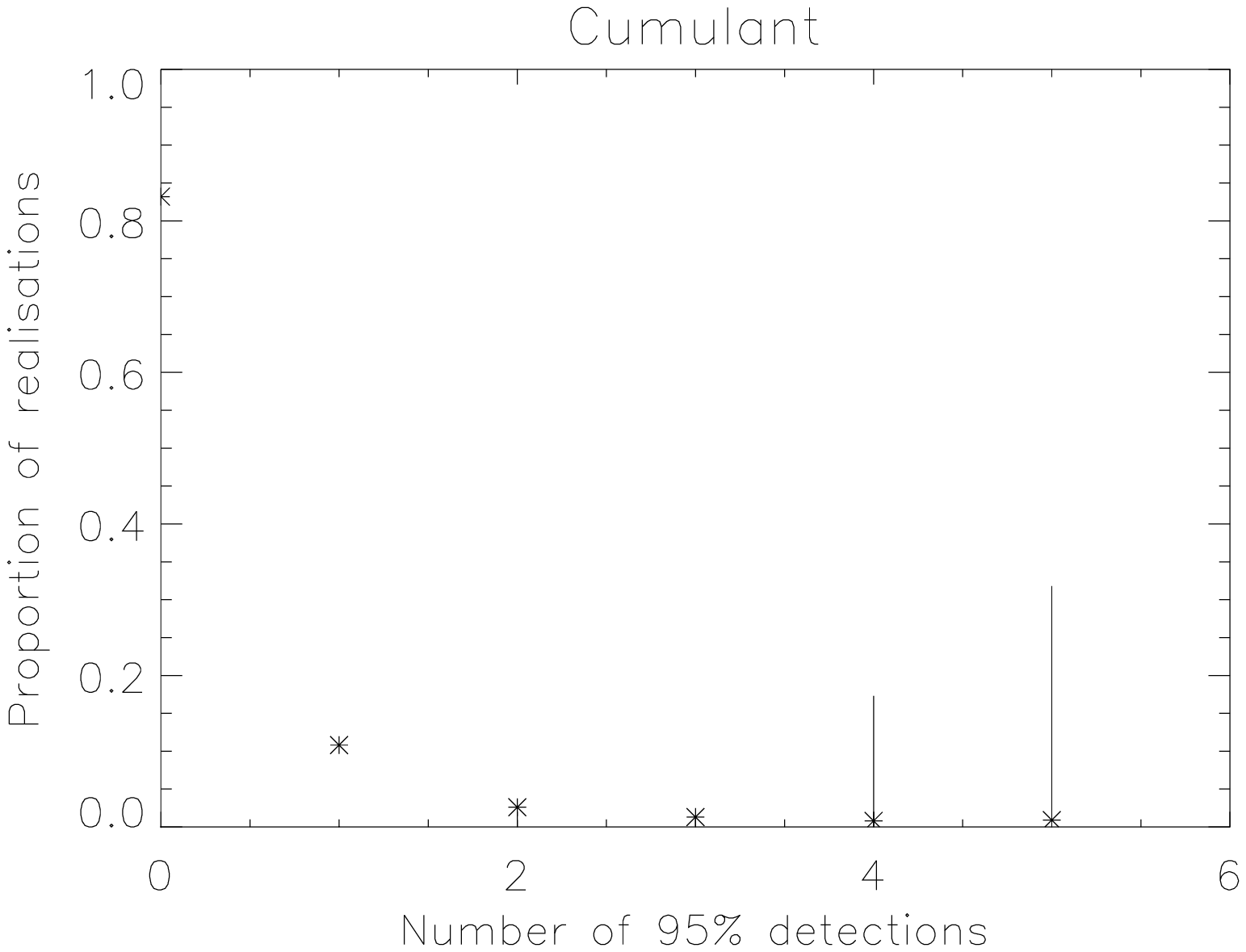 , angle=0, width=5cm}
\epsfig{file=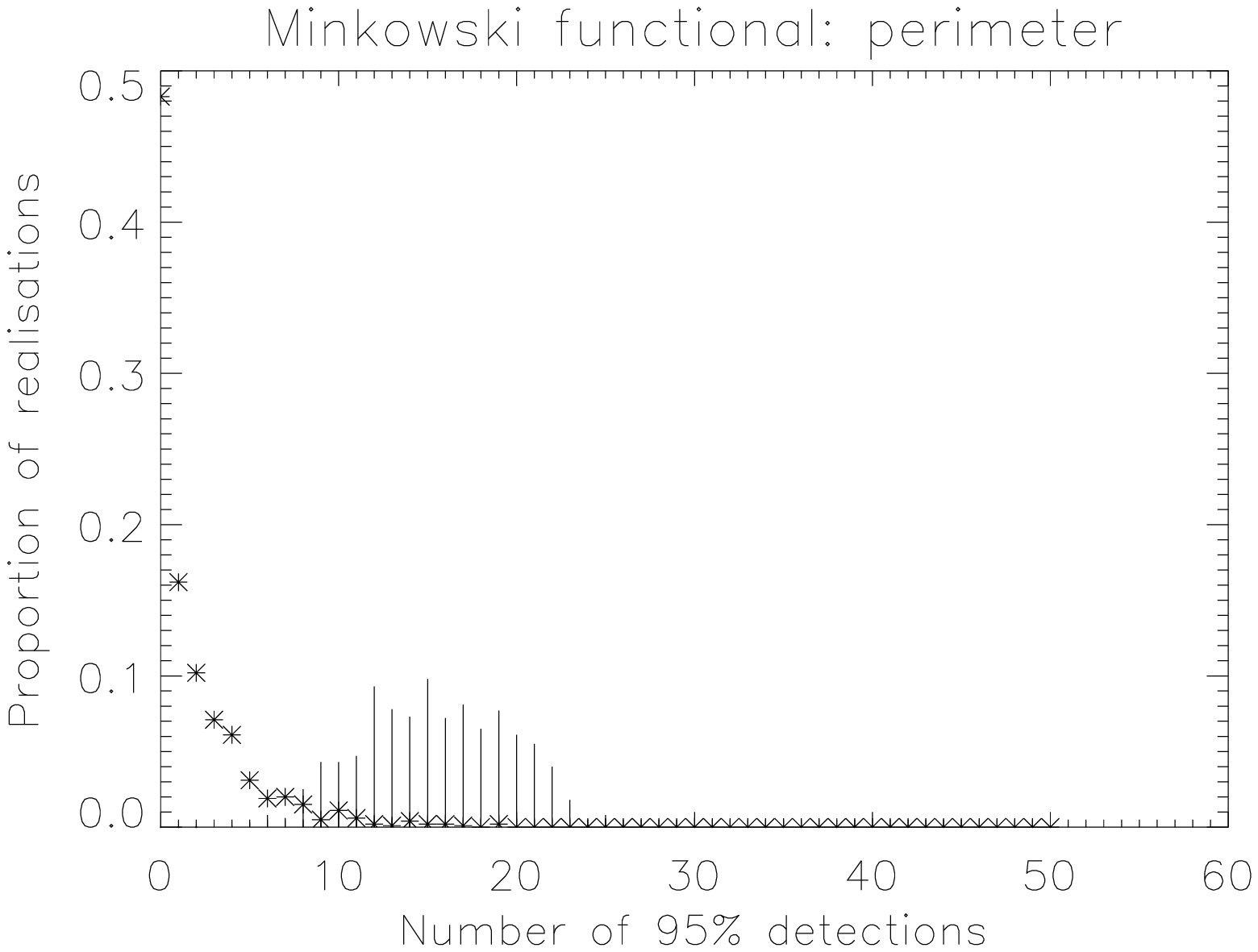 , angle=0, width=5cm}
\epsfig{file=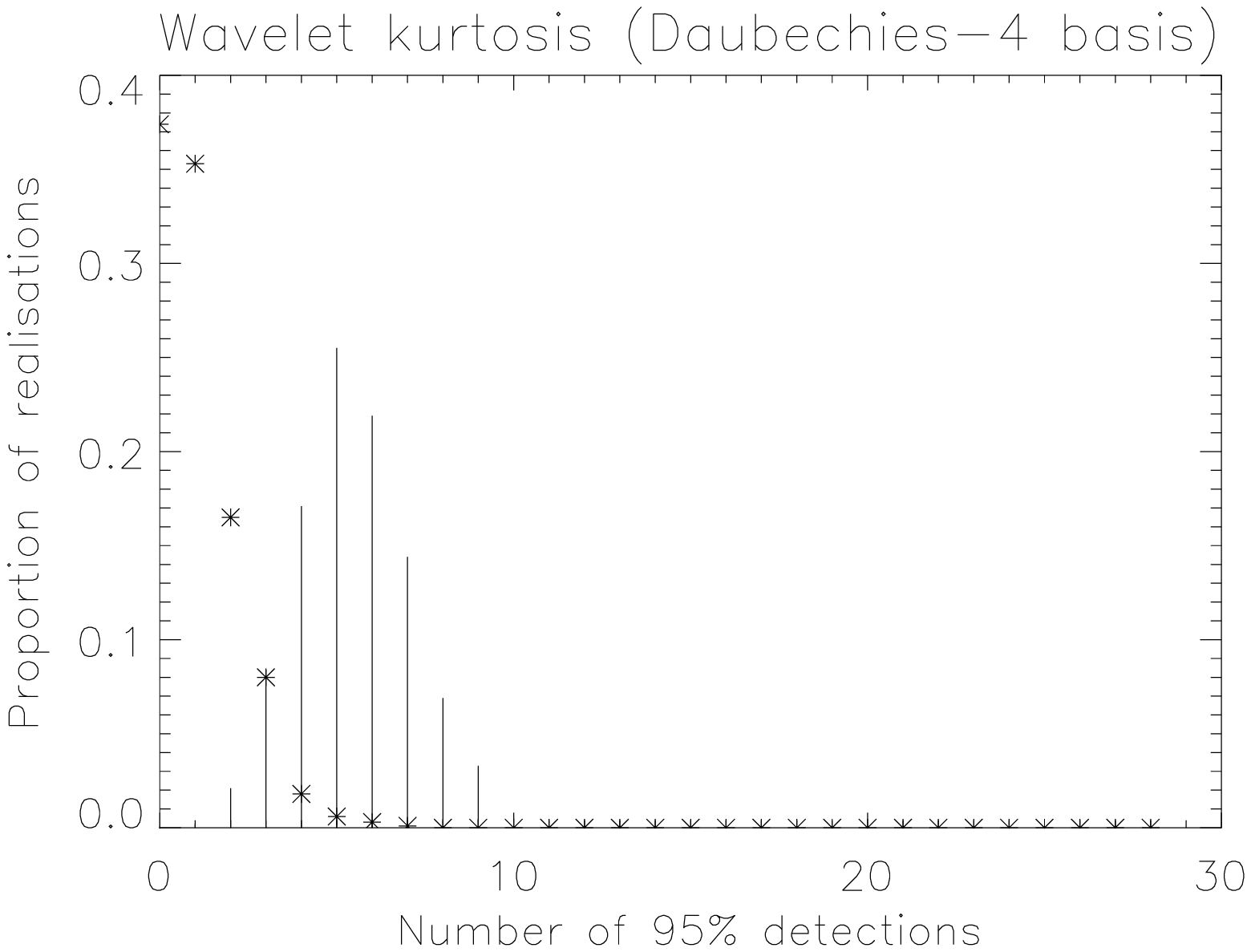 , angle=0, width=5cm}
\end{minipage}
\begin{minipage}{150mm}
\epsfig{file=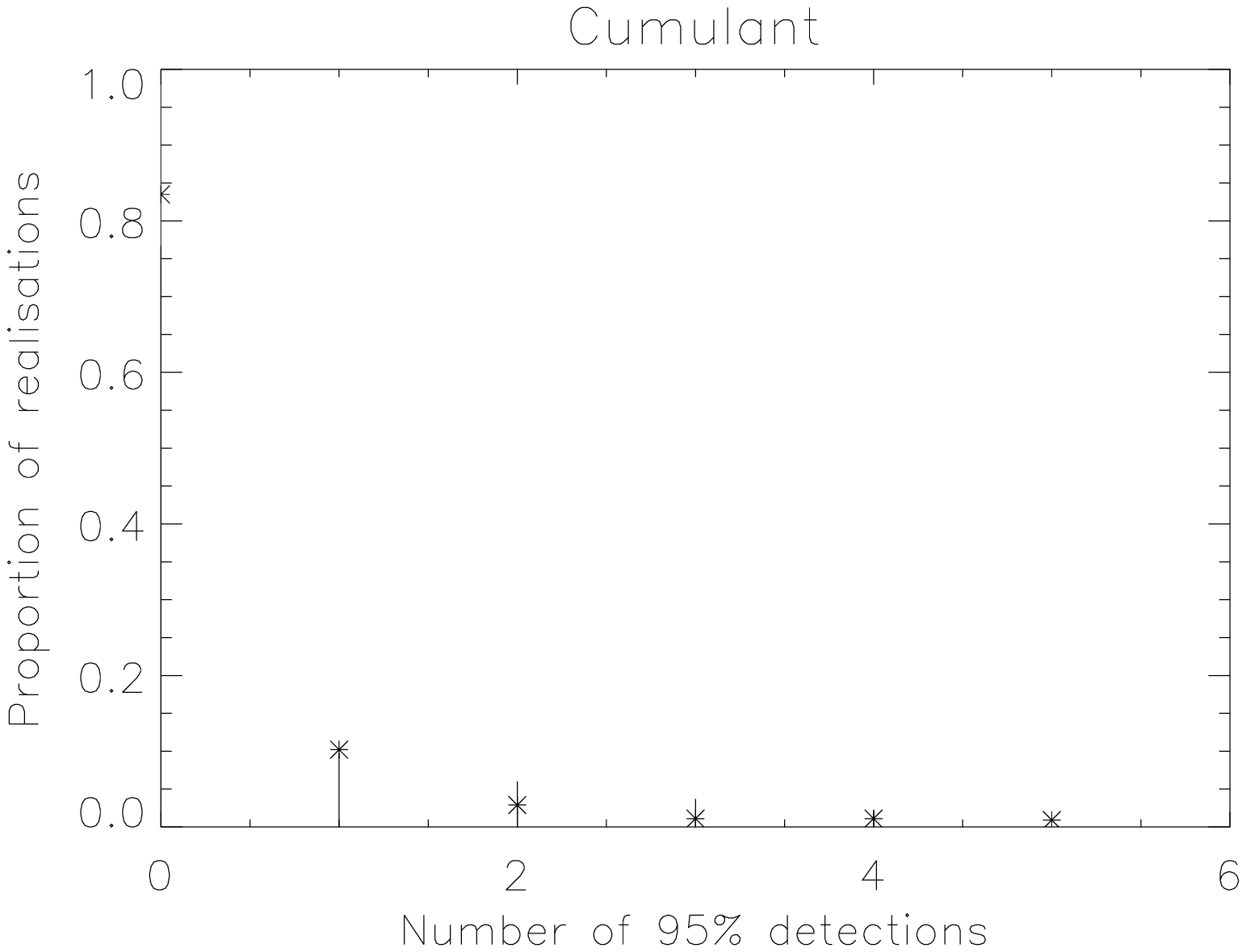 , angle=0, width=5cm}
\epsfig{file=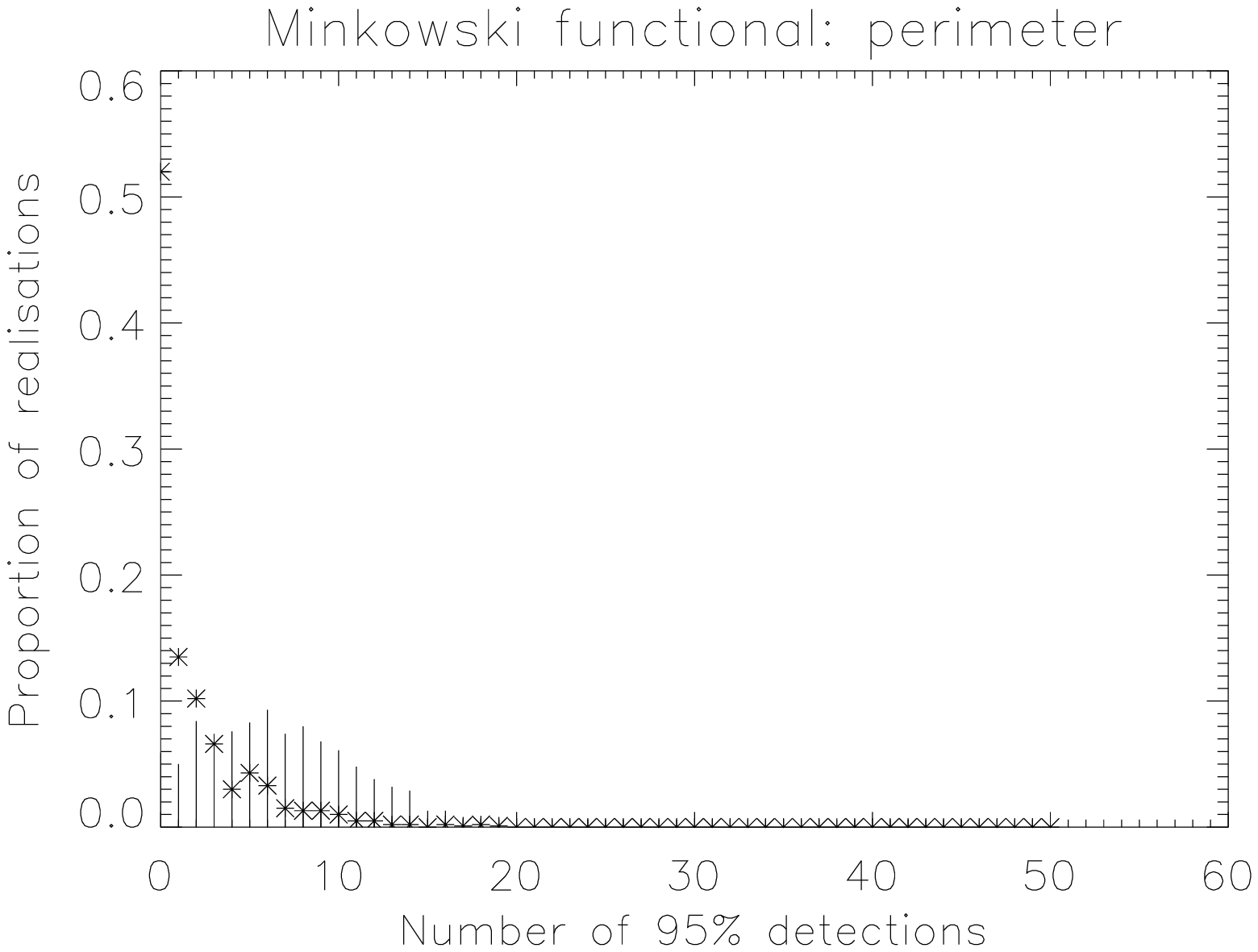 , angle=0, width=5cm}
\epsfig{file=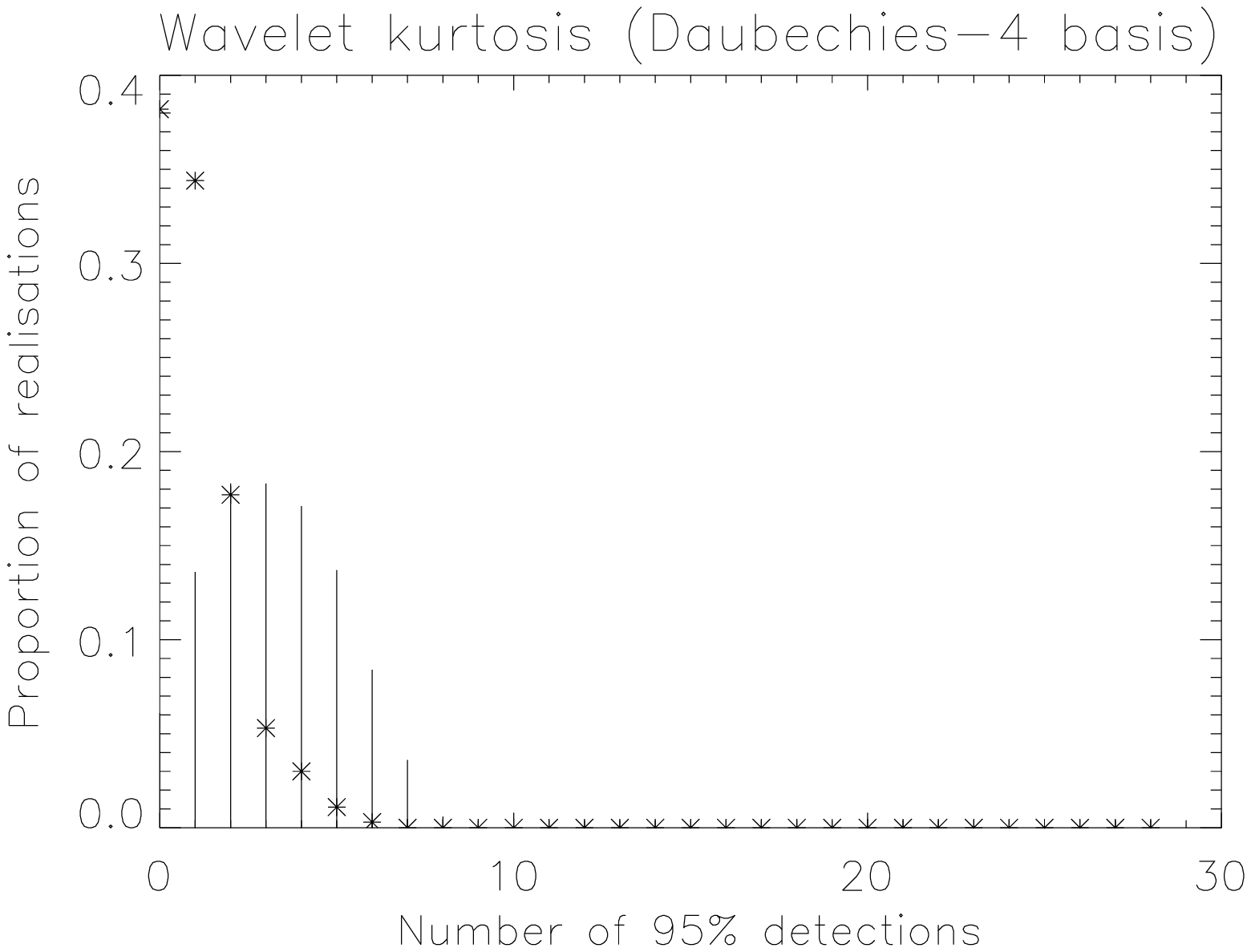 , angle=0, width=5cm}
\end{minipage}
\begin{minipage}{150mm}
\epsfig{file=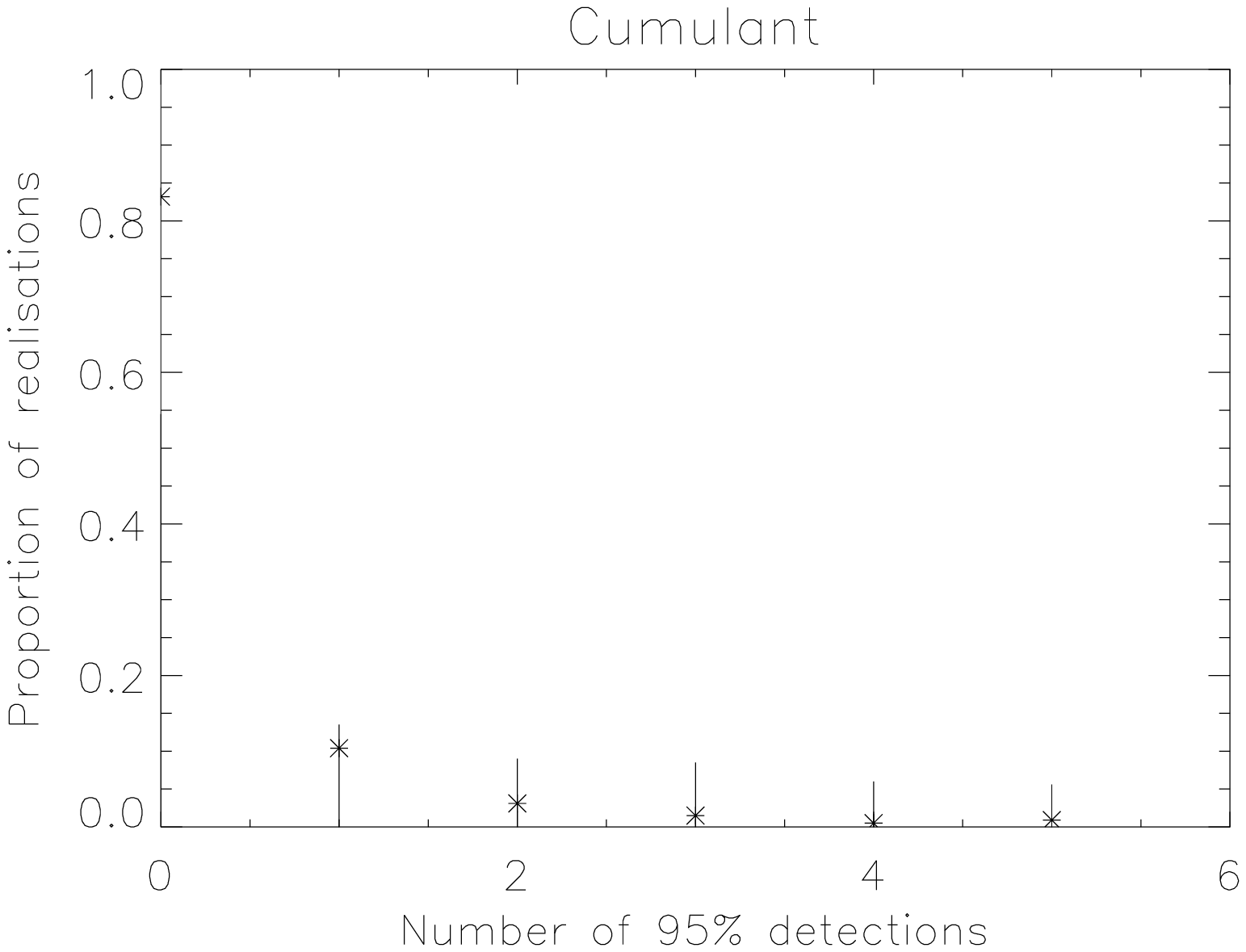 , angle=0, width=5cm}
\epsfig{file=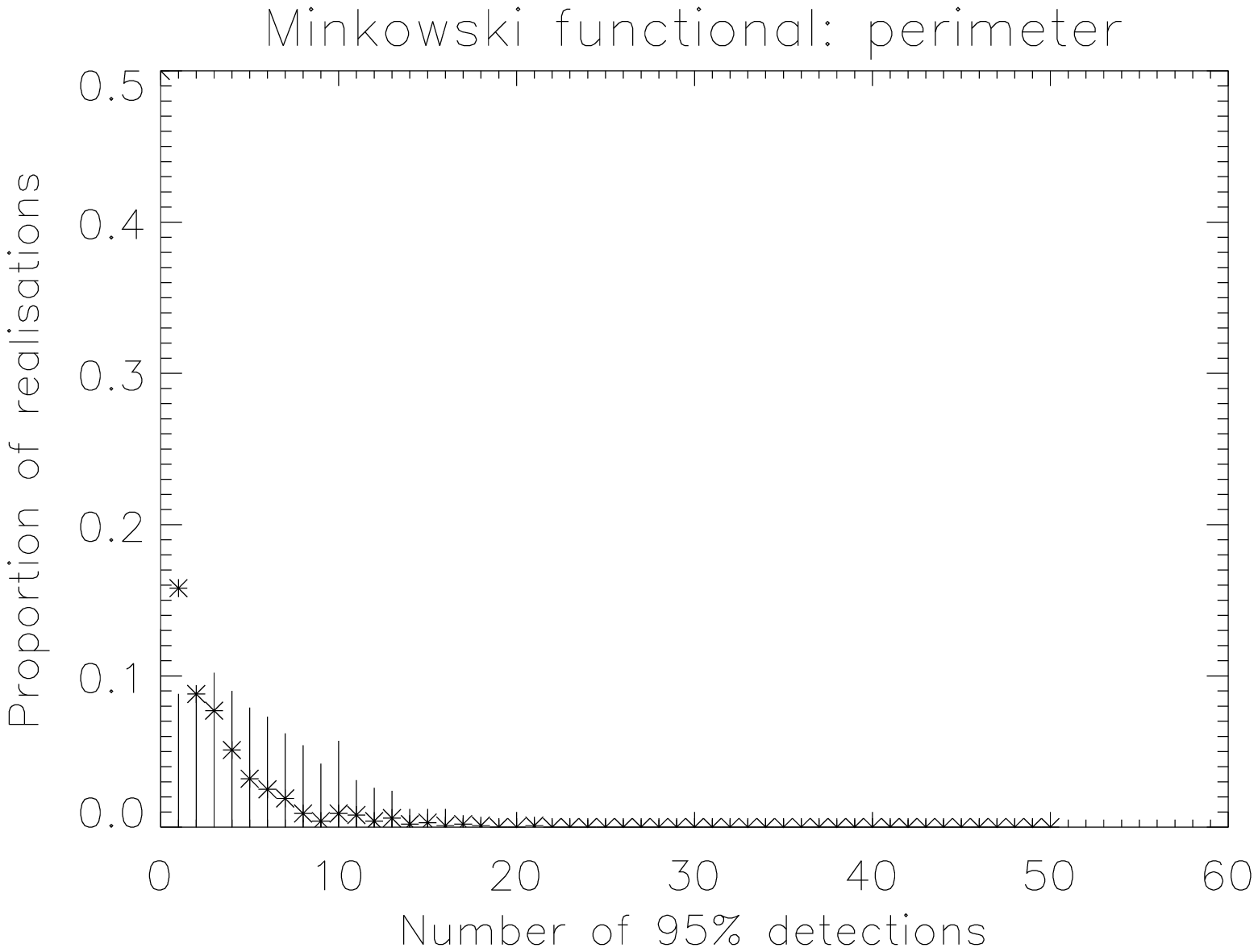 , angle=0, width=5cm}
\epsfig{file=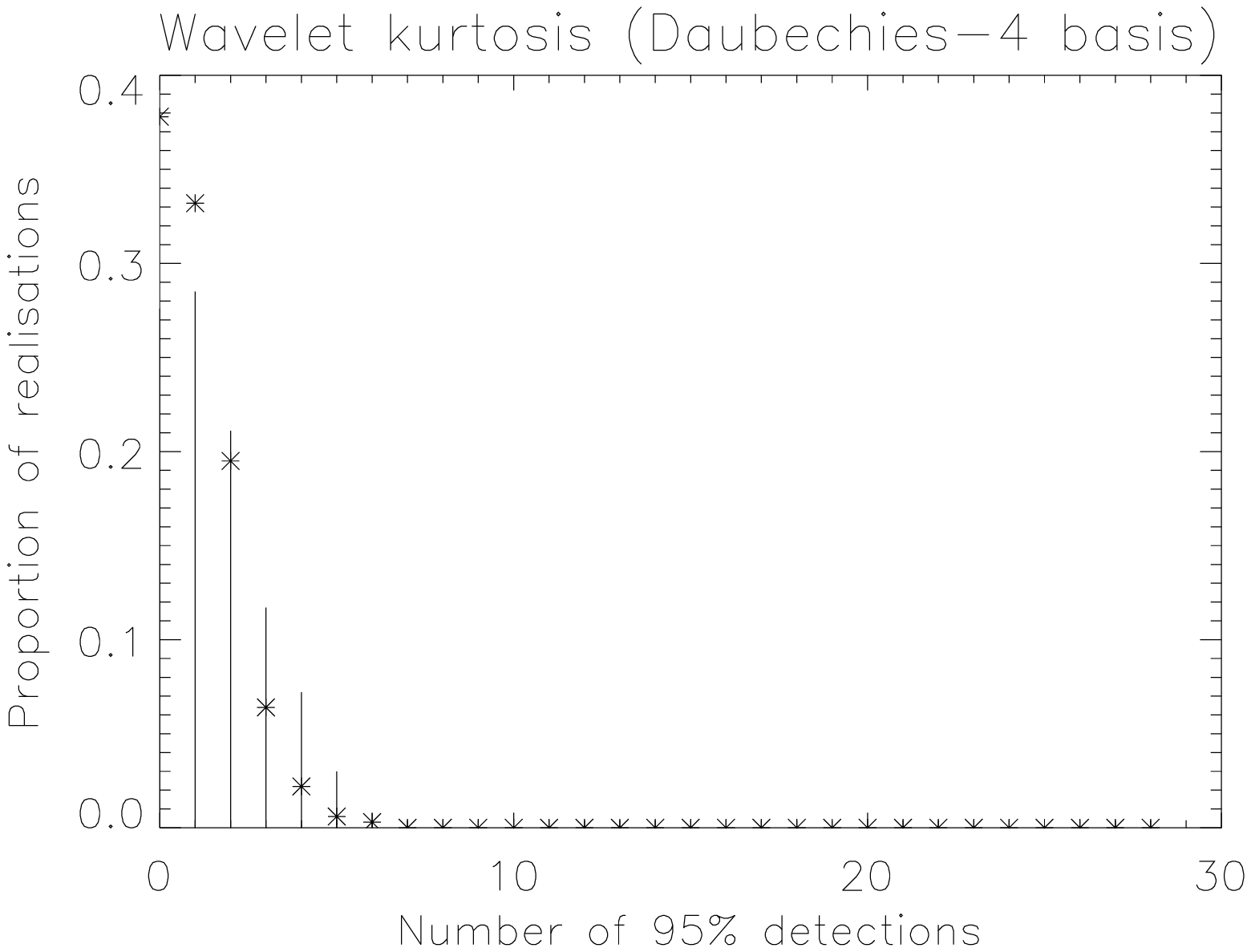 , angle=0, width=5cm}
\end{minipage}
\caption{Example test functions for simulations of various
simulated-skies.  The stars show the test function values for the EGR,
and the vertical bars show the test function values for the
simulations.  Both are calculated from 1000 realisations.
The rows of images show the
results for pure strings (top row), composite strings (second row),
20mJy (third row), 10mJy (fourth row) and 5mJy (5th row)
source subtraction levels.}
\label{fig:sim_test_functions}
\end{figure*}
%
We applied the Bayesian joint estimation analysis to each
of the 6 field centres individually.  Computing constraints mean we
are unable to use Monte Carlo techniques to characterise the
test fully (each analysis takes several hours).  We therefore focus on
unambiguous detections.  

Simulated observations
of the Gaussian CMB field produced no statistically significant
detections.  Moreover,  this approach was found to be insensitive to the
presence of cosmic string networks in both the composite and pure
strings simulations.  However, clear detections were obtained in the
presence of bright point sources.  The most significant detection
occurred in one extended array
field where three sources were present, two with fluxes of order 200~ mJy, and one
of order 100~mJy (see Fig. \ref{fig:hermite_point_source}, solid line).

Analysis of the same simulated data where all point
sources above a flux of 20~mJy (the level of subtraction applied to the
real extended array data) were removed produced a marginal
detection in the same extended array field.  

It is noted that in the visibility plane, information about the sky (in
the form of Fourier modes) has been convolved by the aperture
illumination function.  Any visibility is therefore a linear
combination of several (possibly non-Gaussian) Fourier modes.  This
combination tends, because of the central limit theorem, to lessen the
amount of non-Gaussianity present.  We expect this effect to impair
the ability of the joint estimation analysis to detect cosmological non-Gaussian
structure.  It is, however, ideally suited to detect
non-Gaussianity in the visibility plane which, by a similar argument,
would be poorly detected by map plane statistics.  It is therefore
important to analyse both planes.

\subsection{Map plane simulations}
We made 1000
simulated observations of each sky, each with a
different noise realisation.  By calculating individual statistics and
combining them into test functions for all 1000 simulated observations, as well
as for 1000 EGR, we produce distributions that show the power of any
given statistic or test function to detect the presence of the simulated
sky.  

We applied the map plane analyses to mosaic maps of the simulated
data.  
See Fig. \ref{fig:sim_test_functions} for examples of results of
three of the tests, namely map cumulants, Minkowski functional:
perimeter and the Daubechies-4 kurtosis wavelet test.
We found that simulated observations of the Gaussian CMB field
produced no statistically significant detections in any of our
tests.  

The pure cosmic strings were detected by all three tests.   
In particular, the Minkowski functionals and
wavelet test are clearly highly sensitive to the presence of pure
cosmic strings.  In contrast, none of the tests were sensitive to the
presence of the composite strings/Gaussian CMB simulated-sky.

All of the tests were highly sensitive to the presence of
bright point sources,  and were able to detect the presence of a source distribution up to 20~mJy.  At
10~mJy, the wavelet test is still marginally sensitive to their presence, even
though the other two tests are no longer sufficiently sensitive.  At
5~mJy, all the tests are insensitive to the presence of sources.

\section{Analysis of VSA observations}
We applied all of our chosen non-Gaussianity tests to the VSA
observations.  The joint estimation
test is applied separately to the data for each of the 17 individual
fields.  The map plane tests were applied separately to each of the 3
mosaiced fields.

\subsection{Visibility plane results}
For all 17 individual pointing centres, no statistically significant
deviations from Gaussianity were detected.

\begin{figure}
\epsfig{file=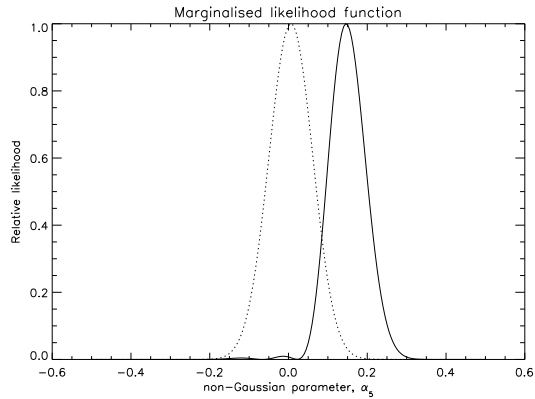 , angle=0, width=7.5cm}
\caption{Example marginalised posterior probability distribution
obtained from analysing one of the compact array fields during data
reduction using the joint estimation analysis.  The data were
subsequently discovered to be contaminated by 
residual spurious signal in a single binned visibility.  The annular bin
analysed here contained 47 binned visibilities.  The solid and dashed
curves show the data with and without the contamination.}
\label{fig:hermite_systematic}
\end{figure}

During the early stages of the reduction of the
compact array fields, this technique detected significant
non-Gaussianity in the data from 3 fields.  Further investigation showed
that these detections were caused by a small number of high amplitude
visibilities in the data (no more than 5 in
each case, and as few as a single point).  In all cases, they were unambiguously
traced to residual spurious signal (see Paper I) and removed.  Fig.
\ref{fig:hermite_systematic} shows an example of the likelihood
functions obtained before and after the removal of a single
contaminated point.

While the basic VSA data reduction procedure has been
demonstrated to deal more than adequately with the spurious signal,
this technique has served as an excellent additional diagnostic of the
VSA data, saving considerable time and effort during data reduction.
It is now used as a standard part of the VSA data reduction process.


%

\subsection{Map plane results}
\begin{figure*}
\begin{minipage}{150mm}
\epsfig{file=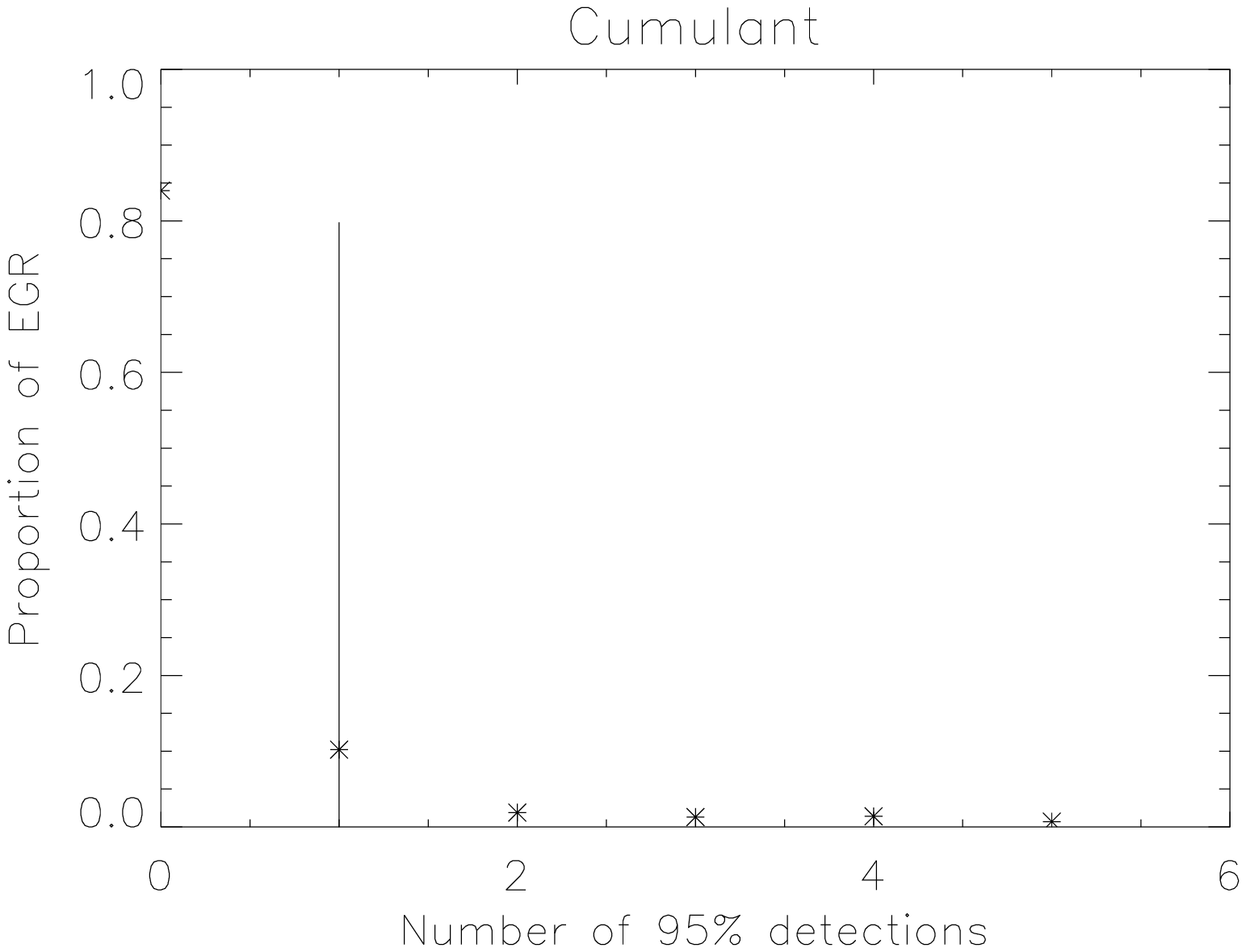 , angle=0, width=5cm}
\epsfig{file=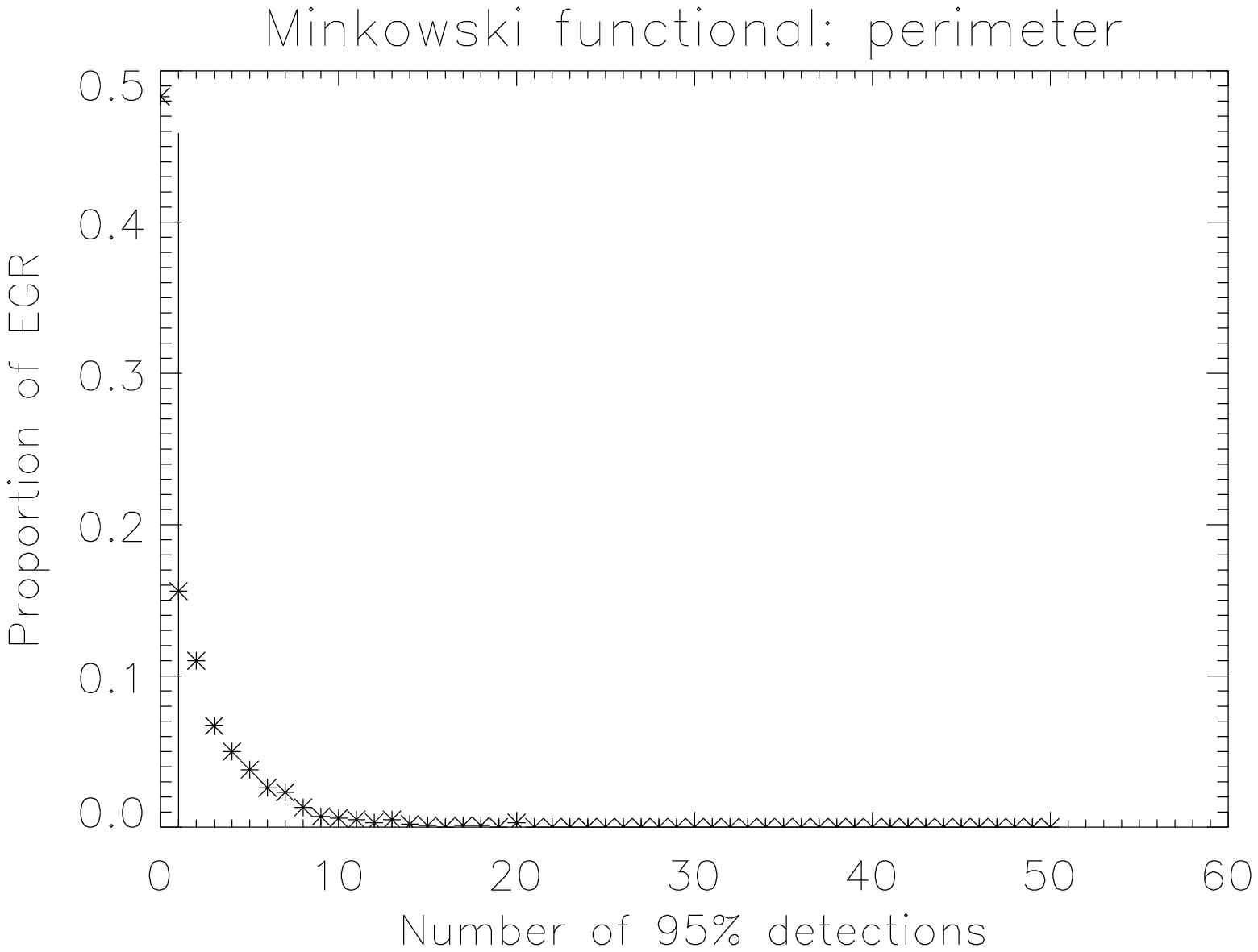 , angle=0, width=5cm}
\epsfig{file=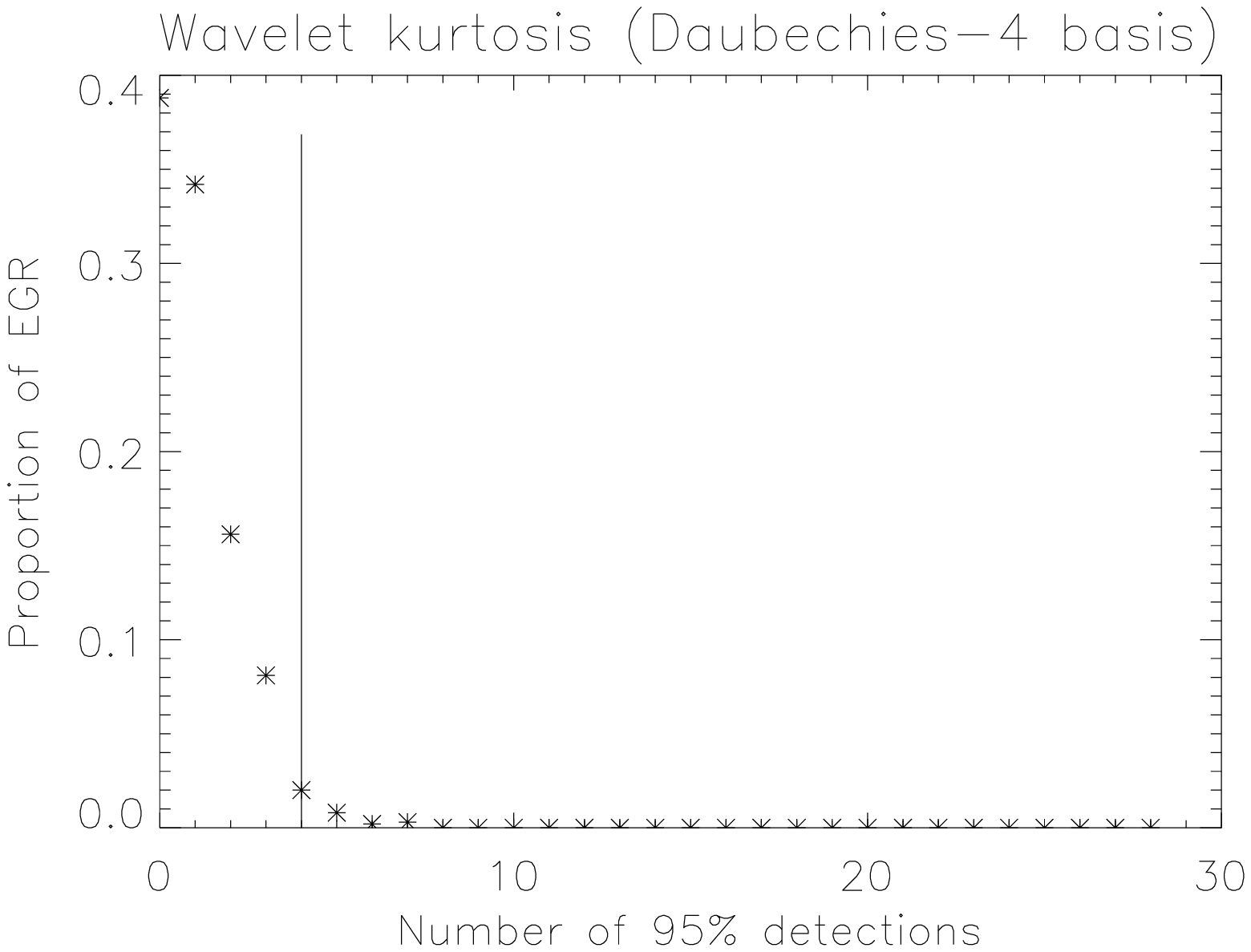 , angle=0, width=5cm}
\caption{VSA1 mosaic example results and Gaussian test functions.  The
Gaussian test functions are denoted by stars, and the value calculated
from the actual data is shown by a vertical bar.}
\label{fig:VSA1_testfn}
\end{minipage}
\begin{minipage}{150mm}
\epsfig{file=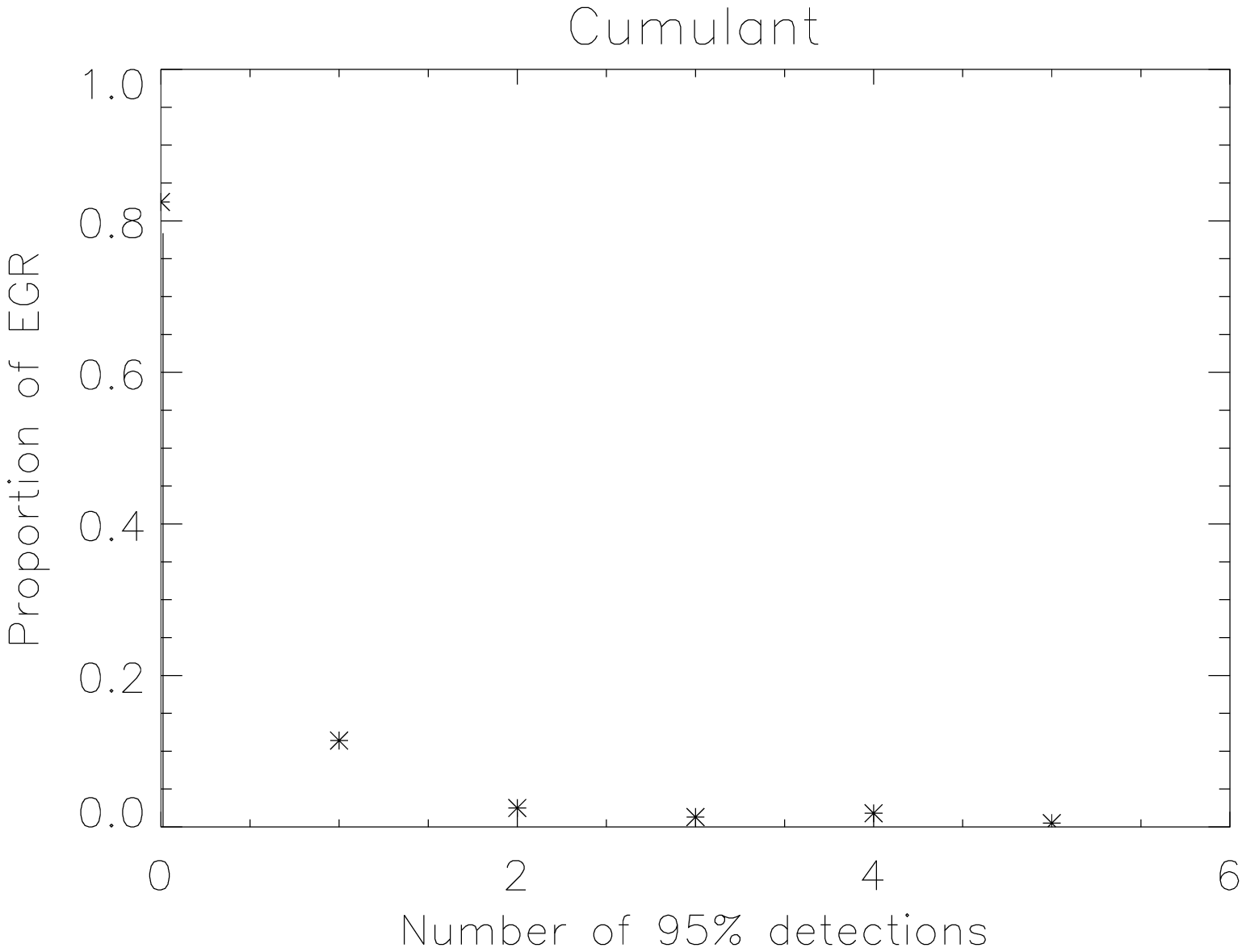 , angle=0, width=5cm}
\epsfig{file=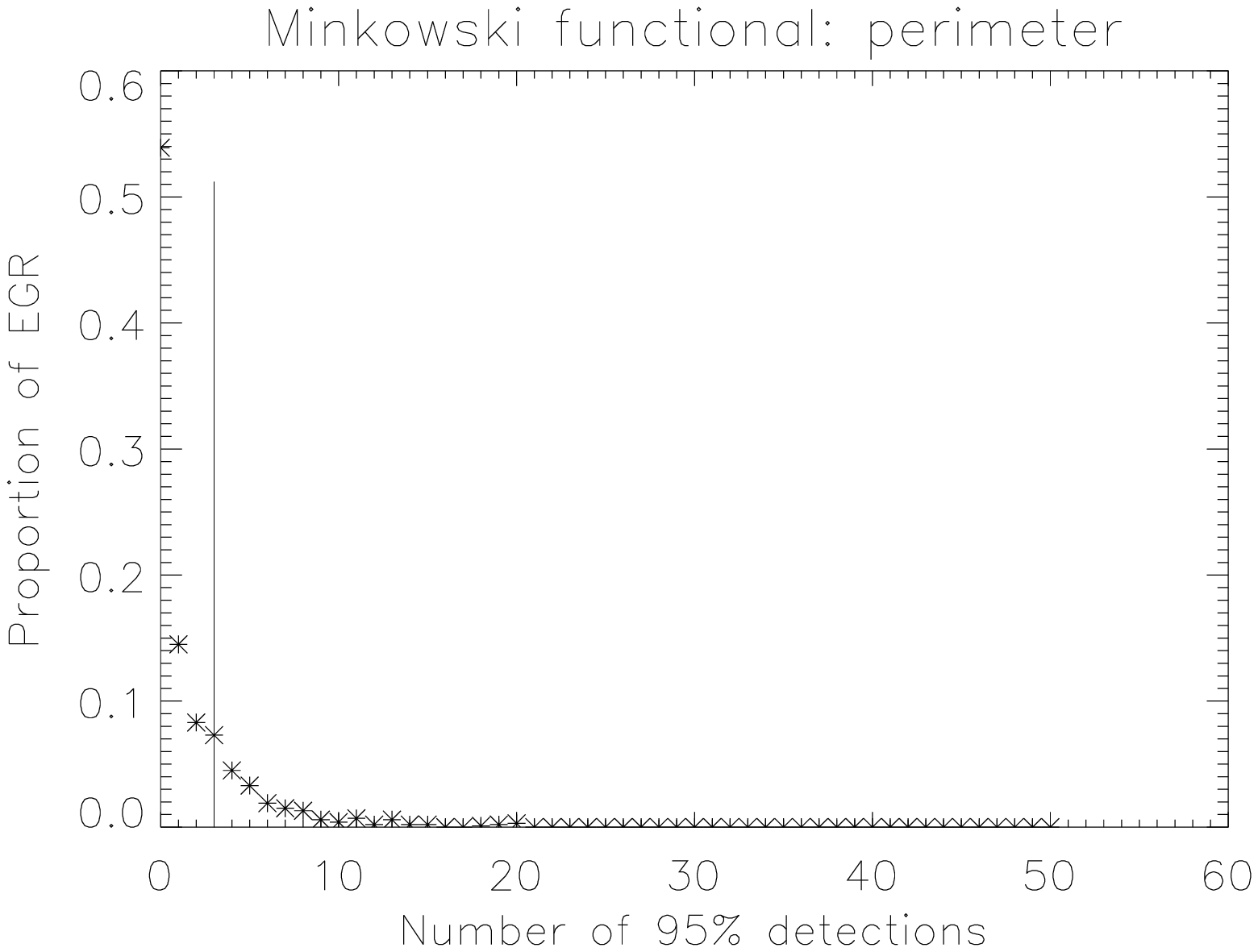 , angle=0, width=5cm}
\epsfig{file=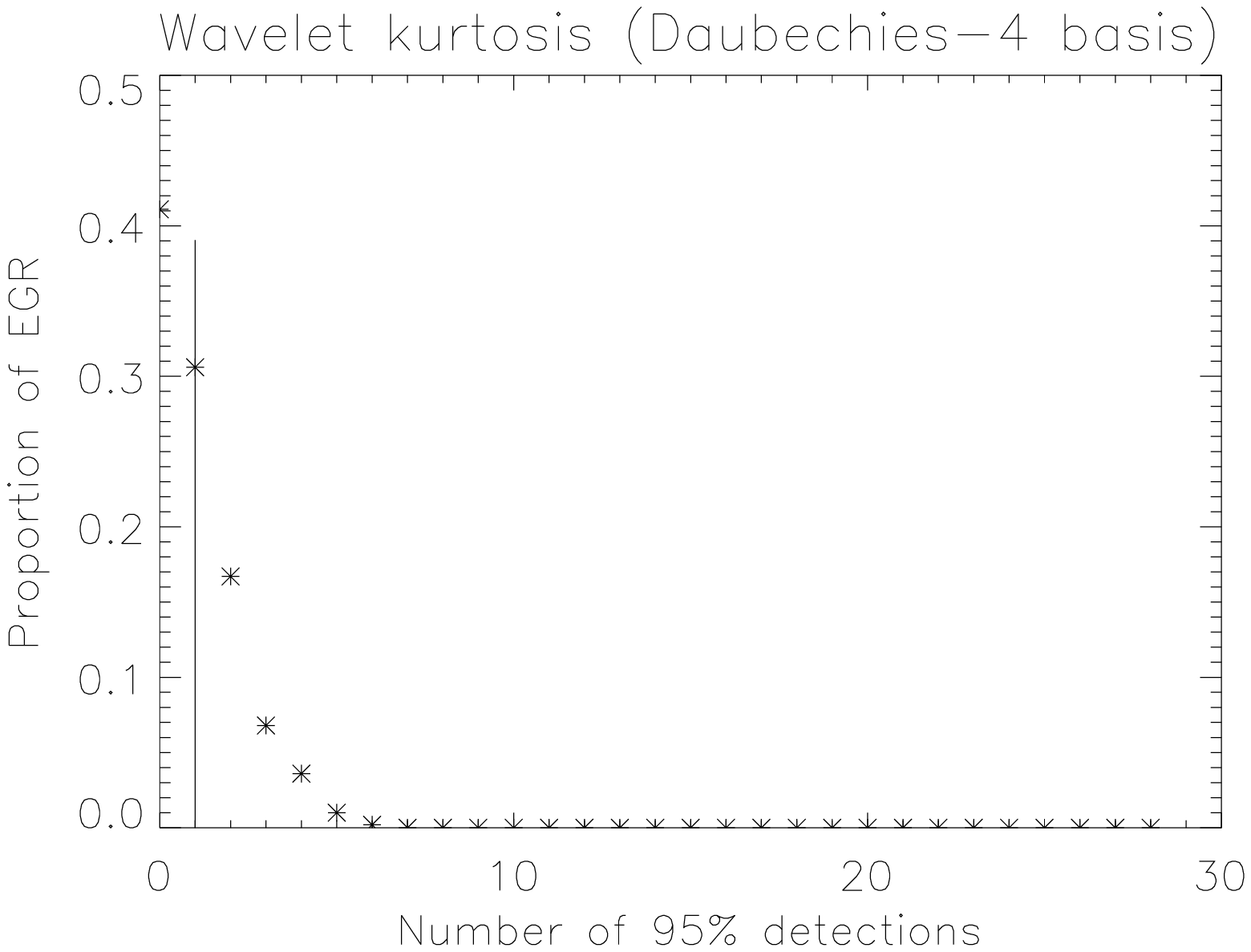 , angle=0, width=5cm}
\caption{VSA2 mosaic example results and Gaussian test functions}
\label{fig:VSA2_testfn}
\end{minipage}
\begin{minipage}{150mm}
\epsfig{file=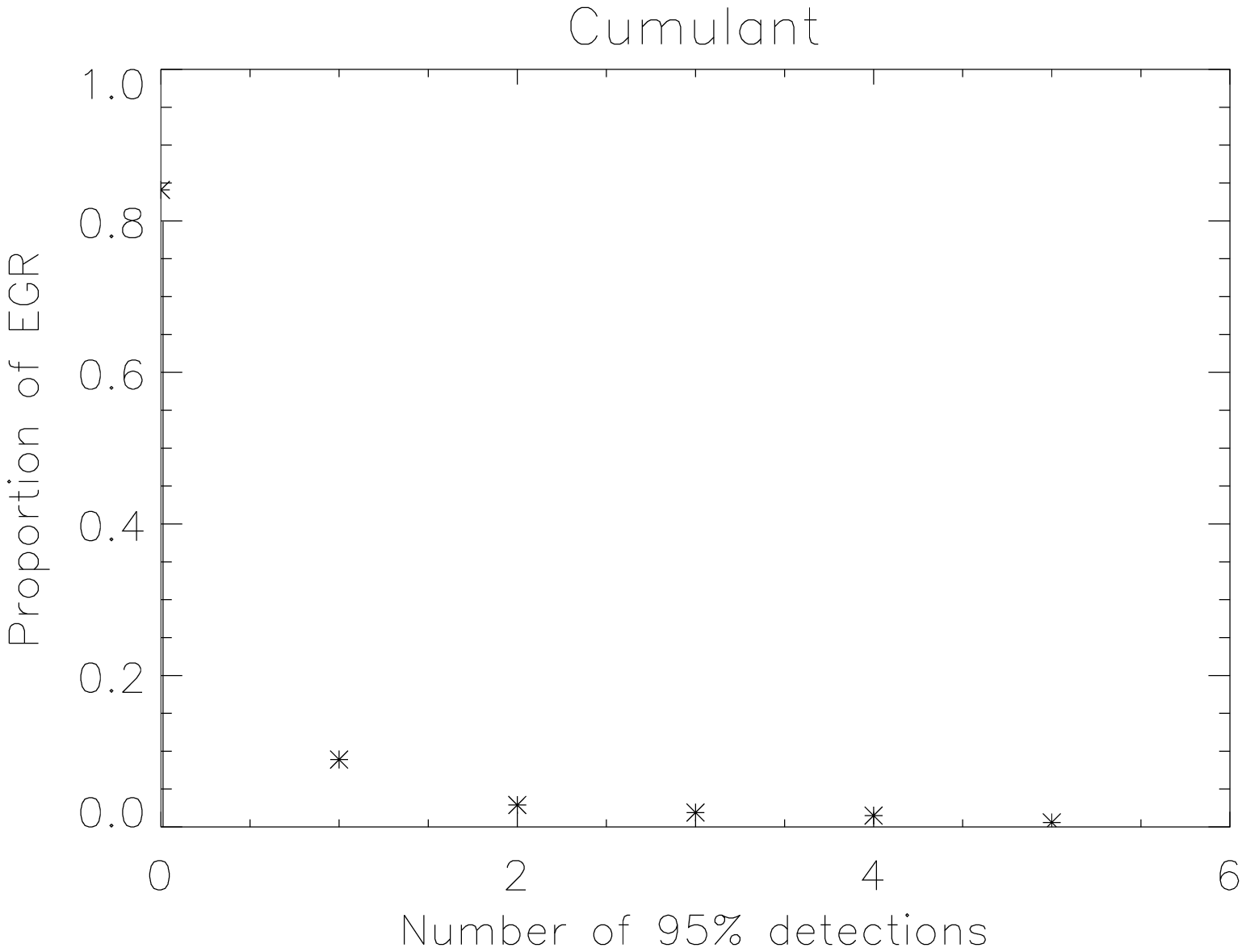 , angle=0, width=5cm}
\epsfig{file=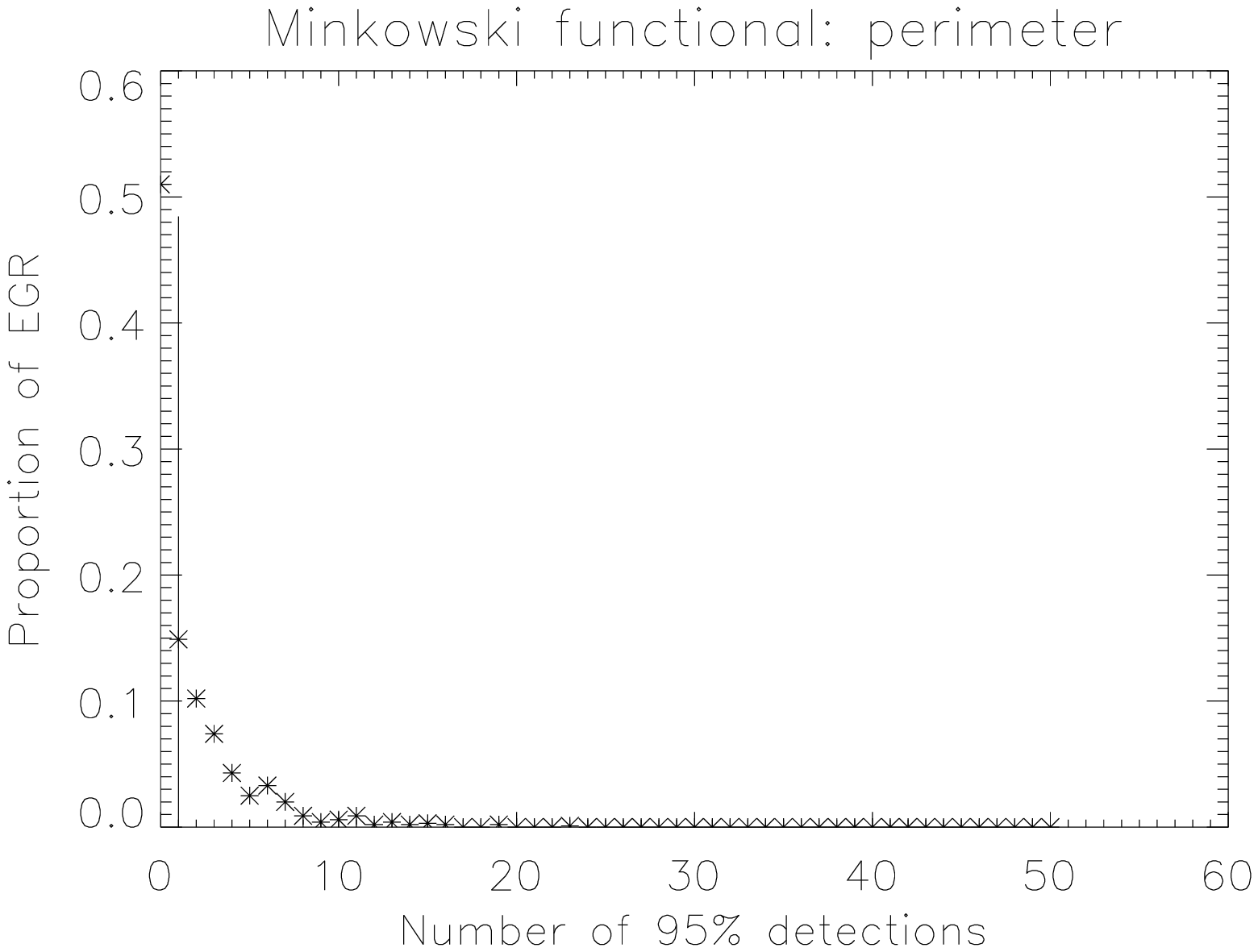 , angle=0, width=5cm}
\epsfig{file=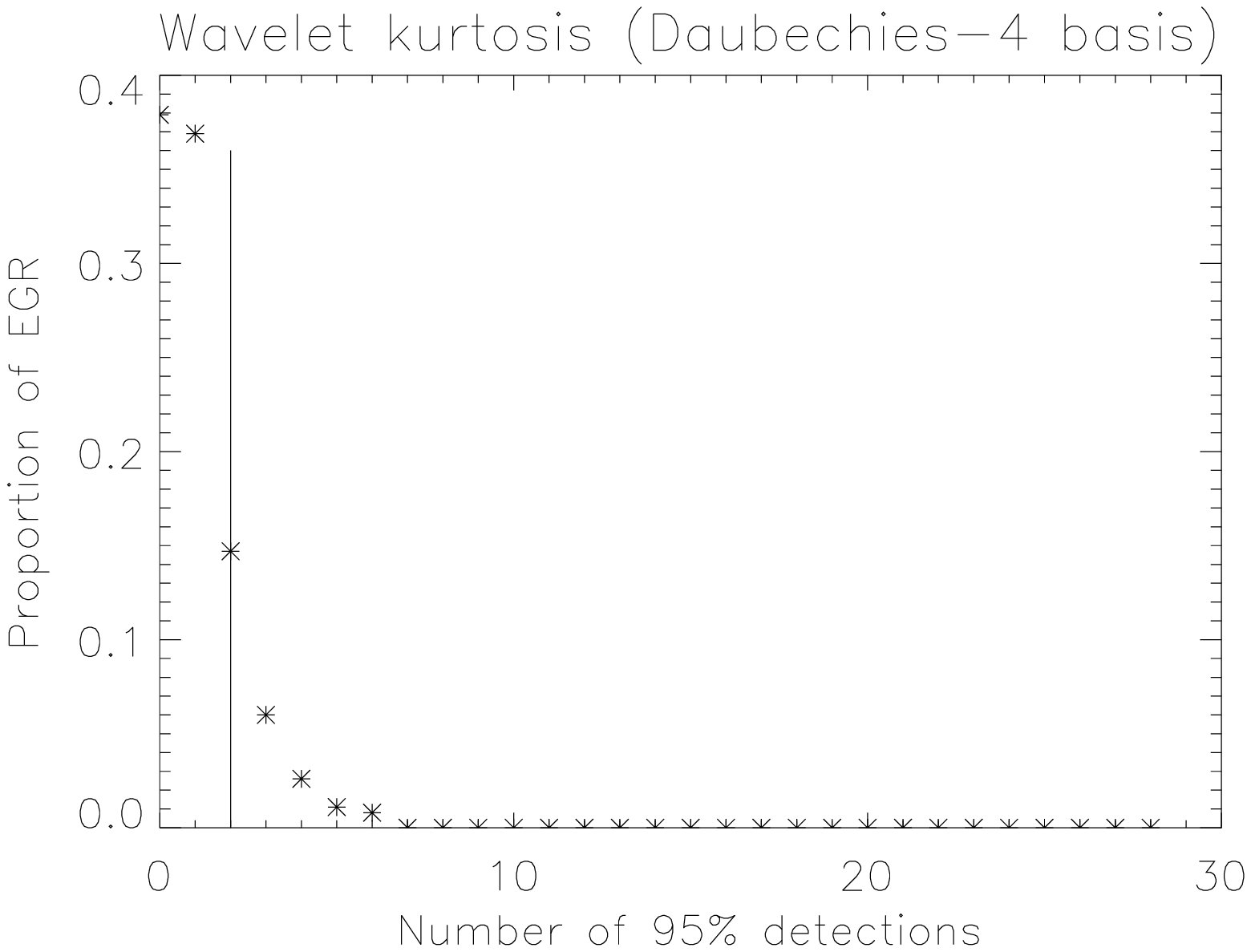 , angle=0, width=5cm}
\caption{VSA3 mosaic example results and Gaussian test functions}
\label{fig:VSA3_testfn}
\end{minipage}
\end{figure*}

The combined test function values and significance levels are given in Table
\ref{tab:map_results}.  The highest significance obtained was 96.7\% ,
in the VSA1 mosaic, from the combined wavelet kurtosis statistics,
using the
Daubechies-4 basis.  The highest significances in the VSA2
and VSA3 mosaics were 87.3\% and 74.8\%, using the area Minkowski
functional and the Daubechies-4 kurtosis wavelet test, respectively.
Figs. \ref{fig:VSA1_testfn}, \ref{fig:VSA2_testfn} and
\ref{fig:VSA3_testfn} show example EGR test functions for the VSA1,
VSA2 and VSA3 mosaics.  In each case, these functions are determined
from the 1000 EGR calculated for each data set. The value of the test
function obtained for the real data is shown as the solid vertical line. 

While the 96.7\% detection in the VSA1 mosaic hints at the presence
of a source of non-Gaussianity, the significance level is not high
enough to rule out Gaussianity.  Moreover, it is
noteworthy that simulations showed the Daubechies-4 kurtosis wavelet test
in particular to be sensitive to realistic point source distributions
below the VSA's source subtraction level.  Therefore, this
detection may be the result of slight point source contamination.
Another point of note is that the VSA1 mosaic is predicted to have
almost double the Galactic foreground contamination of the
other two mosaics  ($5.5~\mu K$, as opposed to $2.7~\mu K$ in the VSA3
mosaic; see Paper II for details).  Whilst this is not a major
contaminant, the Daubechies-4 kurtosis wavelet test may
be sensitive enough to detect it.
The addition of further extended array data to this data set will
significantly improve our sensitivity to non-Gaussianity and may
enable us to either identify the source of the non-Gaussianity in the
VSA1 mosaic, or rule it out as a chance statistical fluctuation.

That significant non-Gaussianity detections are not
found in all three mosaics confirms the success of our source
surveying and subtraction strategy.
Simulations show that if the power law
distribution of radio sources predicted from our source surveying
observations continues below 20~mJy, we would expect a detection 
 in approximately one field in three using the Daubechies-4
kurtosis wavelet test - as indeed we find.  Note
that the VSA power spectrum (see Paper V) includes a statistical correction for
this predicted residual point source contamination. 
\begin{table*}
\begin{minipage}{150mm}
\begin{tabular}{lcccc}
\hline
Case of non-Gaussianity test & number of individual statistics & VSA1 & VSA2 & VSA3 \\
\hline
cumulant 				 & 6  & 1 (84.0\%) & 0 (0\%)    &  0 (0\%) 	 \\
Minkowski: area 			 & 51 & 1 (56.3\%) & 5 (87.3\%) &  2 (69.8\%) \\
Minkowski: perimeter 			 & 51 & 1 (48.3\%) & 3 (76.7\%) &  1 (51.0\%) \\
Minkowski: genus 			 & 51 & 1 (43.3\%) & 2 (70.7\%) &  2 (67.9\%) \\
Wavelet skewness (Haar basis)  		 & 21 & 0 (0\%)    & 1 (43.8\%) &  0 (0\%) 	\\
Wavelet kurtosis (Haar basis) 		 & 21 & 2 (75.7\%) & 3 (87.4\%) &  1 (44.9\%) \\
Wavelet skewness (Daubechies-4 basis)    & 21 & 0 (0\%)    & 1 (39.4\%) &  0 (0\%) 	\\
Wavelet kurtosis (Daubechies-4 basis)    & 21 & 4 (96.7\%) & 1 (41.4\%) &  2 (74.8\%) \\
Wavelet skewness (Daubechies-6 basis)    & 21 & 1 (37.0\%) & 0 (0\%)    &  1 (40.7\%) \\
Wavelet kurtosis (Daubechies-6 basis)    & 21 & 1 (37.0\%) & 0 (0\%)    &  0 (0\%) 	\\
Wavelet skewness (Daubechies-12 basis)   & 21 & 1 (41.3\%) & 0 (0\%)    &  0 (0\%) 	\\
Wavelet kurtosis (Daubechies-12 basis)   & 21 & 3 (91.1\%) & 1 (35.7\%) &  2 (72.6\%) \\
Wavelet skewness (Daubechies-20 basis)   & 21 & 2 (71.3\%) & 1 (40.8\%) &  0 (0\%) 	\\
Wavelet kurtosis (Daubechies-20 basis)   & 21 & 0 (0\%)    & 1 (37.6\%) &  0 (0\%) 	\\
Wavelet skewness (Coiflet-2 basis) 	 & 21 & 1 (42.3\%) & 0 (0\%)    &  1 (41.3\%) \\
Wavelet kurtosis (Coiflet-2 basis) 	 & 21 & 1 (36.0\%) & 0 (0\%)    &  1 (39.2\%) \\
Wavelet skewness (Coiflet-3 basis) 	 & 21 & 2 (72.4\%) & 1 (40.4\%) &  1 (40.5\%) \\
Wavelet kurtosis (Coiflet-3 basis) 	 & 21 & 1 (33.7\%) & 0 (0\%)    &  1 (35.4\%) \\
Wavelet skewness (Symmlet-6 basis) 	 & 21 & 0 (0\%)    & 0 (0\%)    &  0 (0\%) 	\\
Wavelet kurtosis (Symmlet-6 basis) 	 & 21 & 1 (37.0\%) & 1 (37.8\%) &  1 (35.7\%) \\
Wavelet skewness (Symmlet-8 basis) 	 & 21 & 0 (0\%)    & 0 (0\%)    &  0 (0\%) 	\\
Wavelet kurtosis (Symmlet-8 basis) 	 & 21 & 2 (71.0\%) & 1 (36.3\%) &  0 (0\%) 	\\
\hline
Combined Minkowski functionals           &153 & 3 (50.8\%) & 10 (86.0\%)&  5 (68.1\%) \\
Combined Wavelet (Haar basis)            & 42 & 2 (48.4\%) & 4 (80.0\%) &  1 (24.2\%) \\
Combined Wavelet (Daubechies-4 basis)    & 42 & 4 (80.7\%) & 2 (44.8\%) &  2 (45.4\%) \\
Combined Wavelet (Daubechies-6 basis)    & 42 & 2 (45.8\%) & 0 (0\%)    &  1 (19.8\%) \\
Combined Wavelet (Daubechies-12 basis)   & 42 & 4 (82.3\%) & 1 (18.2\%) &  2 (45.7\%) \\
Combined Wavelet (Daubechies-20 basis)   & 42 & 2 (43.1\%) & 2 (45.3\%) &  0 (0\%) \\
Combined Wavelet (Coiflet-2 basis)       & 42 & 2 (45.4\%) & 0 (0\%)    &  2 (45.5\%) \\
Combined Wavelet (Coiflet-3 basis)       & 42 & 3 (67.1\%) & 1 (18.5\%) &  2 (44.3\%) \\
Combined Wavelet (Symmlet-6 basis)       & 42 & 1 (18.2\%) & 1 (19.0\%) &  1 (20.0\%) \\
Combined Wavelet (Symmlet-8 basis)       & 42 & 2 (43.8\%) & 1 (19.6\%) &  0 (0\%) \\
\hline
\end{tabular}
\caption{Test function values and combined significance levels for the
different cases of each map plane non-Gaussianity test.  The results
for each of the three VSA mosaics are given.}
\label{tab:map_results}
\end{minipage}
\end{table*}

\section{Conclusions}
We have analysed the VSA data sets presented in Papers V and VI for
the presence of non-Gaussianity.  We have found the following:
\begin{itemize}
\item{The VSA2 and VSA3 mosaics are consistent with Gaussianity.}
\item{In the VSA1 mosaic, a 96.7\% detection of non-Gaussianity was
found with the kurtosis wavelet test (using the
Daubechies-4 wavelet basis).  We conclude that while this may hint at
a source of non-Gaussianity, the level is consistent with that
expected for the known residual point source contamination in the VSA
data.  It is therefore unlikely to be cosmological in origin.}
\item{Non-Gaussianity testing in the visibility plane
has proven an invaluable tool for locating systematic effects in
data.  It is therefore vitally important to test interferometric observations of the CMB
for non-Gaussianity in both the visibility and map planes.}
\end{itemize}

\section*{ACKNOWLEDGEMENTS}
We thank the staff of the Mullard Radio Astronomy Observatory, Jodrell Bank
Observatory and the Teide Observatory for invaluable assistance in the
commissioning and operation of the VSA. The VSA is supported by PPARC and the
IAC. Partial financial support was provided by Spanish Ministry of Science and
Technology project AYA2001-1657.  R. Savage, K. Lancaster and
N. Rajguru acknowledge  support by PPARC studentships. Pedro Carreira,
K. Cleary and
J. A. Rubi\~no-Martin acknowledge Marie Curie Fellowships of the European
Community programme EARASTARGAL, ``The Evolution of Stars and Galaxies'',
under contract HPMT-CT-2000-00132. K. Maisinger acknowledges support from an
EU Marie Curie Fellowship. A. Slosar acknowledges the support of St. Johns
College, Cambridge. G.Rocha acknowledges support from a Leverhulme
fellowship.  We thank Professor Jasper Wall for assistance and advice
throughout the project, and Anthony Challinor for useful comments.


\label{lastpage}
\bibliography{../ng_refs,../cmb_refs}

\begin{thebibliography}{}

\bibitem[\protect\citeauthoryear{{Aghanim}, {Kunz}, {Castro} \&
  {Forno}}{{Aghanim} et~al.}{2003}]{Aghanim-wavelet-fourier-ng-2003}
{Aghanim} N.,  {Kunz} M.,  {Castro} P.~G.,    {Forno} O.,  2003,
  Non-Gaussianity: Comparing wavelet and Fourier based methods, submitted to
  Astronomy and Astrophysics; astro-ph/0301220

\bibitem[\protect\citeauthoryear{{Banday}, {Zaroubi} \& {G{\' o}rski}}{{Banday}
  et~al.}{2000}]{COBE-systematic}
{Banday} A.~J.,  {Zaroubi} S.,    {G{\' o}rski} K.~M.,  2000, \apj, 533, 575

\bibitem[\protect\citeauthoryear{{Barreiro} \& {Hobson}}{{Barreiro} \&
  {Hobson}}{2001}]{Barreiro-01}
{Barreiro} R.~B.,  {Hobson} M.~P.,  2001, \mnras, 327, 813

\bibitem[\protect\citeauthoryear{{Bartolo} \& {Liddle}}{{Bartolo} \&
  {Liddle}}{2002}]{Bartolo-liddle-curvaton-2002}
{Bartolo} N.,  {Liddle} A.~R.,  2002, \prd, 65, 121301

\bibitem[\protect\citeauthoryear{{Bond}, {Jaffe} \& {Knox}}{{Bond}
  et~al.}{1998}]{bond-98}
{Bond} J.~R.,  {Jaffe} A.~H.,    {Knox} L.,  1998, \prd, 57, 2117

\bibitem[\protect\citeauthoryear{{Bouchet}, {Bennett} \& {Stebbins}}{{Bouchet}
  et~al.}{1988}]{bouchet-strings-1988}
{Bouchet} F.~R.,  {Bennett} D.~P.,    {Stebbins} A.,  1988, \nat, 335, 410

\bibitem[\protect\citeauthoryear{{Chiang}, {Naselsky}, {Verkhodanov} \&
  {Way}}{{Chiang} et~al.}{2003}]{Chiang-wmap-ng-2003}
{Chiang} L.,  {Naselsky} P.~D.,  {Verkhodanov} O.~V.,    {Way} M.~J.,  2003,
  \apjl, 590, L65

\bibitem[\protect\citeauthoryear{{Chiang}, {Naselsky} \& {Coles}}{{Chiang}
  et~al.}{2002}]{Chiang-phase-ng-2002}
{Chiang} L.~Y.,  {Naselsky} P.,    {Coles} P.,  2002, Phase Mapping as a
  Powerful Diagnostic of Primordial Non-Gaussianity, astro-ph/0208235

\bibitem[\protect\citeauthoryear{{Contaldi}, {Bean} \& {Magueijo}}{{Contaldi}
  et~al.}{1999}]{Contaldi-99}
{Contaldi} C.,  {Bean} R.,    {Magueijo} J.,  1999, \prb, 468, 189

\bibitem[\protect\citeauthoryear{{Ferreira}, {Magueijo} \& {Gorski}}{{Ferreira}
  et~al.}{1998}]{Ferreira-COBE-ng-detection-1998}
{Ferreira} P.~G.,  {Magueijo} J.,    {Gorski} K.~M.,  1998, \apjl, 503, L1+

\bibitem[\protect\citeauthoryear{{Ferreira}, {Magueijo} \& {Silk}}{{Ferreira}
  et~al.}{1997}]{Ferreira-cumulants-1997}
{Ferreira} P.~G.,  {Magueijo} J.,    {Silk} J.,  1997, \prd, 56, 4592

\bibitem[\protect\citeauthoryear{{Grainge} et~al.,}{{Grainge}
  et~al.}{2003}]{VSApaperV}
{Grainge} K.,  et~al., 2003, \mnras, 341, L23

\bibitem[\protect\citeauthoryear{{Hansen}, {Marinucci}, {Natoli} \&
  {Vittorio}}{{Hansen} et~al.}{2002}]{Hansen-harmonic-ng-test-2002}
{Hansen} F.~K.,  {Marinucci} D.,  {Natoli} P.,    {Vittorio} N.,  2002, \prd,
  66, 63006

\bibitem[\protect\citeauthoryear{{Hobson}, {Jones} \& {Lasenby}}{{Hobson}
  et~al.}{1999}]{Hobson-wavelet-ng-test}
{Hobson} M.~P.,  {Jones} A.~W.,    {Lasenby} A.~N.,  1999, \mnras, 309, 125

\bibitem[\protect\citeauthoryear{{Hobson} \& {Maisinger}}{{Hobson} \&
  {Maisinger}}{2002}]{madcow}
{Hobson} M.~P.,  {Maisinger} K.,  2002, \mnras, 334, 569

\bibitem[\protect\citeauthoryear{Komatsu et~al.,}{Komatsu
  et~al.}{2003}]{komatsu-wmap-ng-testing-2003}
Komatsu E.,  et~al., 2003, First Year Wilkinson Microwave Anisotropy Probe
  (WMAP) Observations: Tests of Gaussianity, submitted to ApJ; astro-ph/0302223

\bibitem[\protect\citeauthoryear{{Maisinger}, {Hobson} \&
  {Lasenby}}{{Maisinger} et~al.}{1997}]{Maisinger-MEM-1997}
{Maisinger} K.,  {Hobson} M.~P.,    {Lasenby} A.~N.,  1997, \mnras, 290, 313

\bibitem[\protect\citeauthoryear{{Maisinger}, {Hobson}, {Lasenby} \&
  {Turok}}{{Maisinger} et~al.}{1998}]{Maisinger-MEM-topol-1998}
{Maisinger} K.,  {Hobson} M.~P.,  {Lasenby} A.~N.,    {Turok} N.,  1998,
  \mnras, 297, 531

\bibitem[\protect\citeauthoryear{{Mukherjee}, {Hobson} \&
  {Lasenby}}{{Mukherjee} et~al.}{2000}]{Mukherjee-00}
{Mukherjee} P.,  {Hobson} M.~P.,    {Lasenby} A.~N.,  2000, \mnras, 318, 1157

\bibitem[\protect\citeauthoryear{{Peebles}}{{Peebles}}{1999}]{Peebles-ng-infla%
tion-1999}
{Peebles} P.~J.~E.,  1999, \apj, 510, 531

\bibitem[\protect\citeauthoryear{{Polenta} et~al.,}{{Polenta}
  et~al.}{2002}]{Polenta-boomerang-ng-2002}
{Polenta} G.,  et~al., 2002, \apjl, 572, L27

\bibitem[\protect\citeauthoryear{{Rocha}, {Magueijo}, {Hobson} \&
  {Lasenby}}{{Rocha} et~al.}{2001}]{Graca-joint-estimation-2001}
{Rocha} G.,  {Magueijo} J.,  {Hobson} M.,    {Lasenby} A.,  2001, \prd, 64,
  63512

\bibitem[\protect\citeauthoryear{{Rubi{\~ n}o-Martin} et~al.,}{{Rubi{\~
  n}o-Martin}  et~al.}{2003}]{VSApaperIV}
{Rubi{\~ n}o-Martin} J.~A.,  et~al., 2003, \mnras, 341, 1084

\bibitem[\protect\citeauthoryear{{Santos}, {Heavens}, {Balbi}, {Borrill},
  {Ferreira}, {Hanany}, {Jaffe}, {Lee}, {Rabii}, {Richards}, {Smoot},
  {Stompor}, {Winant} \& {Wu}}{{Santos}
  et~al.}{2003}]{Santos-maxima-bispectrum-2003}
{Santos} M.~G.,  {Heavens} A.,  {Balbi} A.,  {Borrill} J.,  {Ferreira} P.~G.,
  {Hanany} S.,  {Jaffe} A.~H.,  {Lee} A.~T.,  {Rabii} B.,  {Richards} P.~L.,
  {Smoot} G.~F.,  {Stompor} R.,  {Winant} C.~D.,    {Wu} J.~H.~P.,  2003,
  \mnras, 341, 623

\bibitem[\protect\citeauthoryear{{Scott} et~al.,}{{Scott}
  et~al.}{2003}]{VSApaperIII}
{Scott} P.~F.,  et~al., 2003, \mnras, 341, 1076

\bibitem[\protect\citeauthoryear{{Slosar} et~al.,}{{Slosar}
  et~al.}{2003}]{VSApaperVI}
{Slosar} A.,  et~al., 2003, \mnras, 341, L29

\bibitem[\protect\citeauthoryear{{Taylor} et~al.,}{{Taylor}
  et~al.}{2003}]{VSApaperII}
{Taylor} A.~C.,  et~al., 2003, \mnras, 341, 1066

\bibitem[\protect\citeauthoryear{Troia et~al.,}{Troia
  et~al.}{2003}]{DeTroia-boomerang-trispectrum-2003}
Troia G.~D.,  et~al., 2003, The trispectrum of the Cosmic Microwave Background
  on sub-degree angular scales: an analysis of the BOOMERanG data, submitted to
  MNRAS; astro-ph/0301294

\bibitem[\protect\citeauthoryear{{Verde} \& {Heavens}}{{Verde} \&
  {Heavens}}{2001}]{Verde-trispectrum-2001}
{Verde} L.,  {Heavens} A.~F.,  2001, \apj, 553, 14

\bibitem[\protect\citeauthoryear{{Waldram}, {Pooley}, {Grainge}, {Jones},
  {Saunders}, {Scott} \& {Taylor}}{{Waldram}
  et~al.}{2003}]{waldram-9C-survey-2003}
{Waldram} E.~M.,  {Pooley} G.~G.,  {Grainge} K.~J.~B.,  {Jones} M.~E.,
  {Saunders} R.~D.~E.,  {Scott} P.~F.,    {Taylor} A.~C.,  2003, \mnras, 342,
  915

\bibitem[\protect\citeauthoryear{{Wang} \& {Kamionkowski}}{{Wang} \&
  {Kamionkowski}}{2000}]{Wang-inflation-bispectrum-2000}
{Wang} L.,  {Kamionkowski} M.,  2000, \prd, 61, 63504

\bibitem[\protect\citeauthoryear{{Watson} et~al.,}{{Watson}
  et~al.}{2003}]{VSApaperI}
{Watson} R.~A.,  et~al., 2003, \mnras, 341, 1057

\bibitem[\protect\citeauthoryear{{Wu}, {Balbi}, {Borrill}, {Ferreira},
  {Hanany}, {Jaffe}, {Lee}, {Rabii}, {Richards}, {Smoot}, {Stompor} \&
  {Winant}}{{Wu} et~al.}{2001}]{Wu-maxima-ng-2001}
{Wu} J.~H.,  {Balbi} A.,  {Borrill} J.,  {Ferreira} P.~G.,  {Hanany} S.,
  {Jaffe} A.~H.,  {Lee} A.~T.,  {Rabii} B.,  {Richards} P.~L.,  {Smoot} G.~F.,
  {Stompor} R.,    {Winant} C.~D.,  2001, Physical Review Letters, 87, 251303

\end{thebibliography}
\bibliographystyle{../mn2e}
\bsp
\end{document}